\documentstyle[preprint,prl,aps]{revtex}
\begin{document} \def\sn2{$\sin^22\theta$} \def\dm2{$\Delta m^2$}
\def\ch2{$\chi^2$} \def\ltap{\ \raisebox{-.4ex}{\rlap{$\sim$}}
\raisebox{.4ex}{$<$}\ } \def\gtap{\ \raisebox{-.4ex}{\rlap{$\sim$}}
\raisebox{.4ex}{$>$}\ } \draft \begin{titlepage}
\preprint{\vbox{\baselineskip 15pt{ \hbox{IASSNS -- AST 95/40}
\hbox{Ref. SISSA 9/95/EP} \hbox{(updated version)}
\hbox{hep-ph/9510367}}}} \vspace*{-0.4cm} \title{\bf
ON THE VACUUM OSCILLATION SOLUTIONS
 OF THE SOLAR NEUTRINO PROBLEM}
\vspace*{-0.4cm}
\author{P. I. Krastev\footnote{\tighten Also at:
Institute of Nuclear Research and Nuclear Energy, Bulgarian Academy
of Sciences, BG--1784 Sofia, Bulgaria.}}
\address{School for Natural Sciences,
Institute for Advanced Study, \\
Princeton, New Jersey 08540,}
\vspace*{-0.4cm}
\author{S.T. Petcov$^*$}
\address{
Scuola Internazionale Superiore di Studi Avanzati, and Istituto\\
Nazionale di Fizica Nucleare, Sezione di Trieste, I-34013 Trieste,
Italy}
\vspace*{-1.0cm}
\maketitle
\begin{abstract}
\begin{minipage}{6.0in}
\baselineskip 15pt
We study the stability of the two--neutrino vacuum oscillation
solution of the solar neutrino problem with respect to changes of the
total fluxes of $^{8}$B and $^{7}$Be neutrinos, $\Phi_{{\rm B}}$ and
$\Phi_{{\rm Be}}$. For any value of $\Phi_{{\rm Be}}$ from the
interval $0.7\Phi^{{\rm BP}}_{{\rm Be}}\leq \Phi_{{\rm Be}}
\leq 1.3\Phi^{{\rm BP}}_{{\rm Be}}$
the solar $\nu_e$ oscillations into an active neutrino, $\nu_e
\leftrightarrow
\nu_{\mu (\tau)}$, provide at 95\% C.L. a description of the existing
solar neutrino data for $\Phi_{{\rm B}} \cong (0.35 - 3.4)
\Phi^{{\rm BP}}_{{\rm B}}$,
$\Phi^{{\rm BP}}_{{\rm B}}$ and $\Phi^{{\rm BP}}_{{\rm Be}}$ being the
fluxes in the solar model of Bahcall--Pinsonneault from 1992. For
$\Phi_{{\rm Be}}\cong (0.7 - 1.3)\Phi^{{\rm BP}}_{{\rm Be}}$ we find
also at 95\% C.L. two new (one new) $\nu_e \leftrightarrow \nu_{\mu
(\tau)}$ ($\nu_e \leftrightarrow
\nu_{s}$) oscillation solutions: i) for
$\Phi_{{\rm B}}\cong (0.35 - 0.43)\Phi^{{\rm BP}}_{{\rm B}}$ at
$\Delta m^2 \cong (4.7 - 6.5)\times 10^{-12}~$eV$^2~((4.8 - 6.4)
\times 10^{-12}~$eV$^2)$ and $\sin^22\theta \gtap 0.71~(0.74)$, and
ii) for $\Phi_{{\rm B}}\cong (0.45 - 0.65)\Phi^{{\rm BP}}_{{\rm B}}$
at $\Delta m^2 \cong (3.2 - 4.0)\times 10^{-11}~$eV$^2$ and
$\sin^22\theta \gtap 0.59$.  The physical implications of the new
solutions for the future solar neutrino experiments are discussed.
The data rule out at 97\% -- 98\% (99 \%) C.L. the possibility of a
universal (neutrino energy independent) suppression of the different
components of the solar neutrino flux, resulting from solar $\nu_e$
oscillations or transitions into active (sterile) neutrino.

\end{minipage}
\end{abstract}
\end{titlepage}
\newpage

\hsize 16.5truecm
\vsize 24.0truecm
\def\dm{$\Delta m^2$\hskip 0.1cm }
\def\dmf{$\Delta m^2$}
\def\sn{$\sin^2 2\theta$\hskip 0.1cm }
\def\snf{$\sin^2 2\theta$}
\def\trna{$\nu_e \rightarrow \nu_a$}
\def\trnm{$\nu_e \rightarrow \nu_{\mu}$}
\def\trns{$\nu_e \leftrightarrow \nu_s$}
\def\trnat{$\nu_e \leftrightarrow \nu_a$}
\def\trnmt{$\nu_e \leftrightarrow \nu_{\mu}$}
\def\trne{$\nu_e \rightarrow \nu_e$}
\def\trnst{$\nu_e \leftrightarrow \nu_s$}
\font\eightrm=cmr8
\def\aprle{\buildrel < \over {_{\sim}}}
\def\aprge{\buildrel > \over {_{\sim}}}
\renewcommand{\thefootnote}{\arabic{footnote}}
\setcounter{footnote}{0}
\vskip 0.0cm
\leftline{\bf 1. INTRODUCTION}
\vskip 0.2cm
\indent In the present paper we investigate the stability of the vacuum
oscillation [1--4] solution of the solar neutrino problem [5,6] with
respect to variations of the total fluxes of the solar $^{8}$B and
$^{7}$Be neutrinos. Recent studies have indicated that the current
solar model predictions [7--12] for the $^{8}$B neutrino flux,
$\Phi_{{\rm B}}$, vary from model to model with rather large
uncertainties [12,13]. The results for $\Phi_{{\rm B}}$ derived in all
solar models presently discussed in the literature except that of
ref. [12], lie in the interval $(4.43 - 6.62)\times
10^{6}~\nu_e$/cm$^{2}$/sec, while the prediction of the "low" flux
model of ref. [12], $\Phi_{{\rm B}} = 2.77 \times
10^{6}~\nu_e$/cm$^{2}$/sec, differs from those of the "high" flux
models of refs. [7] and [11] approximately by the factors 2.0 and
2.4. The predictions [7--12] for the total flux of $^{7}$Be neutrinos,
$\Phi_{{\rm Be}}$, vary by $\sim$20\%, from $\Phi_{{\rm Be}} = 4.34
\times 10^{9}~\nu_e$/cm$^{2}$/sec in ref. [8] to $\Phi_{{\rm Be}} =
5.18 \times 10^{9}~\nu_e$/cm$^{2}$/sec in refs. [11].  At the same
time none of the solar models developed to date provides a
satisfactory description of the existing solar neutrino data
[5,14--16]. In particular, the upper limits on the value of the
$^{7}$Be neutrino flux, which can be inferred from the data, are
considerably lower than the values predicted by the models, as first
noticed in ref. [17] and confirmed in several subsequent more detailed
studies [18] utilizing different methods. The above result follows not
only from joint analyses of the data from all solar neutrino
experiments, Homestake [5], Kamiokande [14], GALLEX [15] and SAGE
[16], but also from the Homestake and Kamiokande, or from the
Kamiokande and SAGE and/or GALLEX data. Since the recent calibration
of the GALLEX detector [19] leaves little room for doubts about the
correctness of the GALLEX results, both the data from the Davis et
al. and Kamiokande experiments have to be incorrect in order for the
indicated conclusion to be not valid.  The discrepancy between the
value of $\Phi_{{\rm Be}}$ suggested by the analyses of the available
solar neutrino data and the solar model predictions for $\Phi_{{\rm
Be}}$ represents a major new aspect of the solar neutrino problem. No
astrophysical and/or nuclear physics explanation of this discrepancy
has been proposed so far.

    Assuming that the $^{7}$Be neutrino flux has a value in the
interval $0.7\Phi^{{\rm BP}}_{{\rm Be}} \leq \Phi_{{\rm Be}}
\leq 1.3\Phi^{{\rm BP}}_{{\rm Be}}$, where
$\Phi^{{\rm BP}}_{{\rm Be}}$ is the flux predicted in the reference
solar model of Bahcall -- Pinsonneault [7], we determine in the
present study the range of values of the $^{8}$B neutrino flux, for
which the results of the solar neutrino experiments can be described
in terms of two--neutrino vacuum oscillations of the solar neutrinos
into an active $\nu_e
\leftrightarrow \nu_{\mu (\tau)}$, or sterile $\nu_e \leftrightarrow
\nu_{s}$, neutrino.
Similar analyses for the MSW solution [20] with solar $\nu_e$
transitions into an active neutrino, $\nu_e \rightarrow
\nu_{\mu (\tau)}$, were performed in refs.  [21,22]. Results for the
case of solar $\nu_e \leftrightarrow \nu_{\mu (\tau)}$ oscillations
were obtained in ref. [23] for $\Phi_{{\rm Be}} = 0.8\Phi^{{\rm
BP}}_{{\rm Be}}$ and $0.4\Phi^{{\rm BP}}_{{\rm B}} \leq
\Phi_{{\rm B}} \leq 1.6\Phi^{{\rm BP}}_{{\rm B}}$, where
$\Phi^{{\rm BP}}_{{\rm B}}$ is the $^{8}$B neutrino flux predicted in
the reference model [7]. However, we find, in particular,
that at 95\% C.L. a new vacuum $\nu_e \leftrightarrow
\nu_{\mu (\tau)}$ ($\nu_e \leftrightarrow \nu_{s}$) oscillation
solution of the solar neutrino problem exists in the region $\Delta
m^2 \cong (4.7 - 6.5)\times 10^{-12}~{\rm eV}^2~((4.8 - 6.4)
\times 10^{-12}~{\rm eV}^2)$ and $0.71~(0.74)
\ltap \sin^22\theta \leq 1.0$
for $\Phi_{{\rm B}}\cong (0.35 - 0.43)\Phi^{{\rm BP}}_{{\rm B}}$ and
$\Phi_{{\rm Be}} = (0.7 - 1.3)\Phi^{{\rm BP}}_{{\rm Be}}$, $\Delta
m^2$ and $\sin^22\theta$ being the two parameters characterizing the
oscillations (see, e.g., refs. [1--4]). A second $\nu_e
\leftrightarrow \nu_{\mu (\tau)}$ oscillation solution is found for
$0.45\Phi^{{\rm BP}}_{{\rm B}} \ltap
\Phi_{{\rm B}} \ltap 0.65\Phi^{{\rm BP}}_{{\rm B}}$ and
$0.7\Phi^{{\rm BP}}_{{\rm Be}} \leq
\Phi_{{\rm Be}} \leq 1.3\Phi^{{\rm BP}}_{{\rm Be}}$, and for values of
$\Delta m^2$ and $\sin^22\theta$ which lie within the intervals
$\Delta m^2 \cong (3.2 - 4.0)\times 10^{-11}~{\rm eV}^2$ and $0.59
\ltap \sin^22\theta \leq 1.0$. Both these solutions were not noticed
in ref. [23]. For $\Delta m^2 > 4.1\times 10^{-11}~{\rm eV}^2$ the
$\nu_e \leftrightarrow \nu_{\mu (\tau)}$ oscillations allow to
describe at 95\% C.L. the existing solar neutrino data for any value
of $\Phi_{{\rm Be}}$ from the interval $(0.7 - 1.3)\Phi^{{\rm
BP}}_{{\rm Be}}$ (for $\Phi_{{\rm Be}} = 0.7\Phi^{{\rm BP}}_{{\rm
Be}}$) and for $0.57~(0.51)\Phi^{{\rm BP}}_{{\rm B}} \ltap \Phi_{{\rm
B}} \ltap 3.4\Phi^{{\rm BP}}_{{\rm B}}$. The corresponding allowed
regions of values of the parameters \dm and \sn in all these cases are
derived as well.  Except for $0.35\Phi^{{\rm BP}}_{{\rm B}} \ltap
\Phi_{{\rm B}} \ltap 0.44\Phi^{{\rm BP}}_{{\rm B}}$ the oscillations
into a sterile neutrino $\nu_e \leftrightarrow \nu_{s}$ are excluded
at 95\% C.L. if $\Phi_{{\rm Be}} = (0.7 - 1.3)\Phi^{{\rm BP}}_{{\rm
Be}}$; at 98\% C.L.  they are allowed by the (mean event rate) solar
neutrino data for $0.32\Phi^{{\rm BP}}_{{\rm B}} \ltap
\Phi_{{\rm B}} \ltap 1.8\Phi^{{\rm BP}}_{{\rm B}}$.

    We have also performed a study which shows that the data do not
favour the hypothesis of neutrino energy independent suppression of
the solar neutrino flux: it is excluded, depending on the value of
$\Phi_{{\rm Be}}$ from the interval $(0.7 - 1.3)\Phi^{{\rm BP}}_{{\rm
Be}}$, at (97\% -- 98\%) C.L. when the suppression is due to $\nu_e
\rightarrow \nu_{\mu (\tau)}$ ($\nu_e \leftrightarrow \nu_{\mu
(\tau)}$) or $\nu_e \rightarrow \bar {\nu}_{\mu (\tau)}$ transitions
(oscillations), and at (99.0\% -- 99.7\%) C.L. if it results from
$\nu_e
\rightarrow \nu_{s}$ ($\nu_e \leftrightarrow \nu_{s}$) transitions
(oscillations).

   The vacuum oscillation solution at $\Delta m^2 \gtap 4.4\times
10^{-11}~{\rm eV}^2~$ imply a non-negligible suppression of the pp
$\nu_e$ flux (approximately by a factor (0.50 -- 0.70)), and a not
very strong suppression of the 0.862 MeV $^{7}$Be $\nu_e$ flux, the
relevant suppression factor ranging from approximately 0.30 (possible
for $\Phi_{{\rm Be}} = 1.3 \Phi^{{\rm BP}}_{{\rm Be}}$) to 0.98
(possible if $\Phi_{{\rm Be}} \cong (0.7 - 1.0) \Phi^{{\rm BP}}_{{\rm
Be}}$ and $\Phi_{{\rm B}} \cong \Phi^{{\rm BP}}_{{\rm B}}$).

    The new $\nu_e \leftrightarrow \nu_{\mu (\tau)}$ ($\nu_e
\leftrightarrow \nu_{s}$) oscillation solution located in the region
$\Delta m^2 \cong (4.7 - 6.5) \times 10^{-12}~{\rm eV}^2~$, which
exists for low values of $\Phi_{{\rm B}}$, is quite similar to the MSW
nonadiabatic ($\nu_e \rightarrow \nu_{\mu (\tau)}$ [21] or $\nu_e
\rightarrow \nu_s$ transition [26]) solutions for similar values of
$\Phi_{{\rm B}}$: it corresponds to a rather strong suppression of the
0.862 MeV $^{7}$Be $\nu_e$ flux (by a factor (0.06 -- 0.26)), to a
moderate suppression of the pp neutrino flux (by a factor not less
than 0.7 only) and to $^{8}$B neutrino flux practically not affected
by the oscillations.  The physical implications of the indicated new
vacuum oscillation solutions for the future solar neutrino experiments
are also briefly discussed.  Our results show, in particular, that the
$\nu_e \leftrightarrow \nu_{\mu (\tau)}$ vacuum oscillation solution
of the solar neutrino problem is stable with respect to changes of the
predictions for the $^{8}$B and $^{7}$Be neutrino fluxes.

   We use the latest published data from all four solar neutrino
experiments [5,14--16] in our analysis: $$\bar{\rm R}({\rm Ar}) =
(2.55~\pm~0.25)\hskip 0.2cm {\rm SNU},~~~\eqno(1)$$
$$\bar\Phi^{{\rm exp}}_{{\rm B}} = (2.89~\pm~0.42) \times 10^{6}~{\rm
cm^{-2}sec^{-1}},~~\eqno(2)$$ $$\bar {\rm R}_{\rm GALLEX}({\rm Ge}) =
(77.1~\pm~9.9)\hskip 0.2cm {\rm SNU},~~\eqno(3)$$ $$\bar {\rm R}_{{\rm
SAGE}}({\rm Ge}) = (69~\pm~13)\hskip 0.2cm {\rm SNU},~~\eqno (4)$$
\noindent where
$\bar{\rm R}$(Ar), and $\bar{\rm R}_{\rm GALLEX}({\rm Ge})$ and
$\bar{\rm R}_{\rm SAGE}({\rm Ge})$, are respectively the average rates
of $^{37}$Ar and $^{71}$Ge production by solar neutrinos observed in
the experiments of Davis et al. [5], and GALLEX [15] and SAGE [16],
and $\bar{\Phi}^{{\rm exp}}_{{\rm B}}$ is the flux of $^{8}$B
neutrinos measured by the Kamiokande collaborations [14]. In eqs. (1)
-- (4) the quoted errors represent the added in quadratures
statistical (1 s.d.) and systematical errors.
\vskip 0.2cm
\leftline{\bf 2. THE $^{8}$B AND $^{7}$Be NEUTRINO FLUXES}
\vskip 0.1cm
   It is convenient to introduce the parameters $${\rm f_{B}} \equiv
{{\rm \Phi_{B}}\over {\rm \Phi_{B}^{BP}}} \geq 0,~~~~~ {\rm f_{Be}}
\equiv {{\rm \Phi_{Be}}\over {\rm \Phi_{Be}^{BP}}} \geq 0,
{}~~~\eqno(5)$$
\noindent in terms of which we shall describe the possible
deviations of ${\rm \Phi_{B}}$ and ${\rm \Phi_{Be}}$ from their values
in the reference model [7]. The fluxes ${\rm \Phi_{B}}$ and ${\rm
\Phi_{Be}}$ in the models [7,8,11,12] correspond, respectively, to
$~{\rm f_{B}}=~$ 1.0; 0.78; 1.16; 0.49, and $~{\rm f_{Be}}=~$1.0;
0.89; 1.06; 0.89.

    The Kamiokande data, evidently, imposes limits on the values ${\rm
\Phi_{B}}$ (and f$_{{\rm B}}$) can possibly have.  The expression for
the predicted event rate in the Kamiokande detector, R(K), if the
$^{8}$B (electron) neutrinos undergo two--neutrino transitions into an
active neutrino $\nu_{\mu (\tau)}$ (due to vacuum oscillations
$\nu_{e} \leftrightarrow \nu_{\mu (\tau)}$ or MSW transitions $\nu_{e}
\rightarrow \nu_{\mu (\tau)}$), or $\bar {\nu}_{\mu (\tau)}$ (due to
spin--flavour conversion $\nu_{e} \rightarrow \bar {\nu}_{\mu
(\tau)}$) on their way to the Earth, has the form \footnote{\tighten
In the numerical calculations we have performed we have included the
Kamiokande energy resolution and trigger efficiency functions in the
expression under the integral in eq. (6).}: $${\rm R(K)} = {\rm
f_{B}\Phi_{B}^{BP}}\int\limits_{}^{14.4~{\rm MeV}}~ {\rm
n(E)~\sigma_{K}(E)~[P(E)}~+~0.16(1 - {\rm P(E))]}~d{\rm E},~~ \eqno
(6)$$
\noindent where n(E)
is the normalized to 1 spectrum of $^{8}$B neutrinos,
$\int\limits_{}^{14.4~{\rm MeV}}{\rm n(E)}d{\rm E} = 1$, ${\rm
\sigma_{K}(E)}$ is the $\nu_e - {\rm e}^{-}$ elastic scattering
cross--section for $^{8}$B neutrinos with energy E, in which the
recoil e$^{-}$ detection efficiency and energy resolution functions of
the Kamiokande detector are included, P(E) is the probability of
survival of the $^{8}$B $\nu_e$ having energy E ((1 -- P(E)) is the
probability of the $\nu_{e} \rightarrow \nu_{\mu (\tau)}$ transition
due to vacuum oscillations or the MSW effect, or of the $\nu_{e}
\rightarrow \bar {\nu}_{\mu (\tau)}$ conversion), and we have used the
fact that $\sigma_{\nu_{\mu (\tau)}e}$(E)/$\sigma_{\nu_{e}e}$(E)~$
\cong
\sigma_{\bar {\nu}_{\mu
(\tau)}e}$(E)/$\sigma_{\nu_{e}e}$(E)$~\cong 0.16$ in the energy range
of interest, $\sigma_{\nu_{l} e}$(E) and $\sigma_{\bar {\nu}_{l}
e}$(E), l=e,$\mu,\tau$, being the $\nu_{l} - {\rm e}^{-}$ and $\bar
{\nu}_{l} - {\rm e}^{-}$ elastic scattering cross--sections. In the
case of $\nu_{e} \leftrightarrow \nu_{s}$ oscillations or $\nu_{e}
\rightarrow \nu_{s}$ transitions the term with the coefficient 0.16 is
absent from the expression in the right hand side of eq.  (6).

     Given R(K), ${\rm \Phi_{B}^{BP}}$, n(E) and ${\rm
\sigma_{K}(E)}$, the minimal allowed value of ${\rm f_{B}}$, as it
follows from (6), is determined by the maximal possible value of [P(E)
+ 0.16 (1 -- P(E))], which is 1 and is reached when P(E) = 1.  Thus,
we have ${\rm f_{B}} \geq {\rm R(K)}$/${\rm R_{BP}(K)} =
(0.51~\pm~0.07)$, where ${\rm R_{BP}(K)}$ is the event rate predicted
in the BP model [7], and we have used the Kamiokande result, eq. (2).
At 99.73\% (95\%) C.L. this implies $${\rm f_{B}} \gtap
0.30~(0.37).~~~~~\eqno (7)$$

   It is trivial to convince oneself that the above lower limit on
${\rm f_{B}}$ holds also in the case of solar $\nu_e$ two--neutrino
oscillations (transitions) into sterile neutrino $\nu_s$, as well as
for oscillations (transitions) involving more than two neutrinos
(sterile and/or active).  The limit (7) is universal: it does not
depend on the type of possible oscillations (transitions), and on the
specific mechanism responsible for them.

    Similarly, the maximal allowed value of ${\rm f_{B}}$ by the
Kamiokande data corresponds to min [P(E) + 0.16 (1 -- P(E))] =
0.16. We have then: ${\rm f_{B}} \leq {\rm R(K)}$/$(0.16~{\rm
R_{BP}(K)}) = (3.2~\pm~0.44)$, which gives at 99.73\% (95\%) C.L.
$${\rm f_{B}} \ltap 4.5~(4.1).~~~~~\eqno (8)$$
\noindent   Inequality (8) is universal for two--neutrino solar
$\nu_e$ oscillations or transitions into an active neutrino $\nu_{\mu
(\tau)}$ or $\bar {\nu}_{\mu (\tau)}$.

    Contrary to the lower limit (7), the upper limit (8) is not valid
for two--neutrino $\nu_{e} \leftrightarrow \nu_{s}$ ($\nu_{e}
\rightarrow \nu_{s}$)
oscillations (transitions) or $\nu_e$ oscillations (transitions)
involving more than two neutrinos. In the first case, for instance,
the maximal value of $~{\rm f_{B}}$ would correspond to the min P(E),
and the use of the general property of the probability P(E), min P(E)
= 0, does not allow one to derive a useful upper limit on ${\rm
f_{B}}$ from the Kamiokande data.

     In our study of the stability of the results on the vacuum
oscillation solutions with respect to ${\rm \Phi_{B}}$ and ${\rm
\Phi_{Be}}$ variations the following approach is adopted. The fluxes
of the pp, pep and the CNO neutrinos (see, e.g., refs. [6]) are kept
fixed and their values were taken from ref. [7]. The fluxes of the
$^{8}$B and $^{7}$Be neutrinos, and correspondingly, f$_{{\rm B}}$ and
f$_{{\rm Be}}$, are treated as fixed parameters, which, however, are
allowed to take any values within certain intervals. In the case of
${\rm \Phi_{Be}}$ the interval chosen corresponds to $$0.7 \leq {\rm
f}_{{\rm Be}} \leq 1.3.~~~~\eqno(9)$$
\noindent It is somewhat wider than the interval formed by the current
solar model predictions: 0.89 -- 1.06. For ${\rm \Phi_{B}}$ values in
the intervals determined by the inequalities (7) and (8) were
considered. The searches for a $\nu_{e} \leftrightarrow \nu_{s}$
oscillation solution were preformed for $0.3 \leq {\rm f}_{{\rm B}}
\leq 4.0$.

    The indicated approach was motivated by the fact that the
contributions of the CNO neutrinos to the signals in all three types
of detectors [5,14--16] are predicted to be relatively small [6--12],
and that (apart from the CNO neutrinos) the spreads in the predictions
for the fluxes ${\rm \Phi_{B}}$ and ${\rm \Phi_{Be}}$ are the
largest. Some of the values of ${\rm \Phi_{Be}}$ used in the analyses,
as those corresponding to f$_{{\rm Be}}=~$0.7 and 1.3, for example,
are incompatible with the constraint on the solar neutrino fluxes
which the data on the solar luminosity impose (see, e.g.,
refs. [27,28]): $$\Phi_{\rm pp} + 0.958\Phi_{\rm Be} + 0.955 \Phi_{\rm
CNO} + 0.910\Phi_{\rm pep}= (6.517~\pm~0.02) \times 10^{10}~{\rm
cm}^{-2}{\rm sec}^{-1},~~\eqno (10)$$
\noindent where $\Phi_{\rm CNO} = \Phi_{\rm N} + \Phi_{\rm O}$,
and $\Phi_{\rm pp}$, $\Phi_{\rm pep}$, $\Phi_{\rm N}$ and $\Phi_{\rm
O}$ are the fluxes of the pp, pep and the CNO neutrinos.  However, a
20\% -- 30\% change in ${\rm \Phi_{Be}}$ with respect to ${\rm
\Phi_{Be}^{BP}} = 4.89 \times 10^{9}~\nu_e$/cm$^{2}$/sec is required
by (10) to be balanced by only a few percent change of the pp neutrino
flux, and the latter will have a small effect on the predictions for
the signal in the Ga--Ge experiments [15,16]. Besides, the aim of the
present study (as like of the analogous studies of the MSW solutions
in refs. [21,22,26]) is, in particular, to determine the ranges of
values of ${\rm \Phi_{B}}$ and ${\rm \Phi_{Be}}$ for which the
possibility of vacuum oscillations of solar neutrinos cannot be
excluded by the existing solar neutrino data.  Certainly, values of
${\rm \Phi_{B}}$ corresponding to, e.g., f$_{{\rm B}} \cong~$3 seem at
present unlikely to appear in any realistic solar model.

   In the absence of "unconventional behaviour" (vacuum oscillations,
MSW transitions, etc.) of solar neutrinos, the signals in the Cl--Ar
and Ga--Ge experiments can be written in the following form within the
above approach: $${\rm R(Ar)} = (6.20{\rm f_{B}} + 1.17{\rm f_{Be}} +
0.40_{\rm CNO} + 0.23_{\rm pep})~{\rm SNU},~~\eqno (11)$$ $${\rm
R(Ge)} = (70.8_{\rm pp} + 3.1_{\rm pep} + 35.8{\rm f_{Be}} + 13.8{\rm
f_{B}} + 7.9_{\rm CNO})~{\rm SNU},~~\eqno (12)$$
\noindent where 6.20${\rm f_{B}~SNU}$ is the contribution in R(Ar) due
to the $^{8}$B neutrinos, etc.

   We have used the $\chi^2-$method in our analysis. In computing the
$\chi^2$ for a given pair of values of the parameters $\Delta m^2$ and
$\sin^22\theta$ we have ignored the estimated uncertainties in the
reference model predictions [7] for the solar neutrino fluxes as the
ranges within which we have varied ${\rm \Phi_{B}}$ and ${\rm
\Phi_{Be}}$ exceed by far the uncertainties.  We did, however, take
into account the uncertainties in the detection cross--sections for
the detectors [5,14--16].
\vskip 0.3cm
\leftline{\bf 3. THE VACUUM OSCILLATION SOLUTIONS}
\vskip 0.2cm
\leftline{
\bf 3.1 The Case of $\nu_{e} \leftrightarrow \nu_{\mu (\tau)}$
Oscillations}
\vskip 0.2cm
\leftline{\bf 3.1.1 Allowed Regions of the Parameters}
\vskip 0.1cm
\indent Searching for vacuum $\nu_{e} \leftrightarrow \nu_{\mu (\tau)}$
and $\nu_{e} \leftrightarrow \nu_{s}$ oscillation solutions we have
scanned the region $10^{-12}~{\rm eV}^2 \leq \Delta m^2 \leq
10^{-9}~{\rm eV}^2$ and $10^{-2} \leq \sin^22\theta \leq 1.0$. It was
found that at 95\% C.L.  and for $0.7 \leq {\rm f}_{{\rm Be}} \leq
1.3$ the two--neutrino $\nu_{e} \leftrightarrow \nu_{\mu (\tau)}$
oscillations of the solar $\nu_e$ allow one to describe the data (1)
-- (4) for rather large intervals of values of ${\rm f}_{{\rm
B}}$. These intervals depend somewhat on the value of ${\rm f}_{{\rm
Be}}$. Below we give the solution intervals for ${\rm f}_{{\rm B}}$
(at 95\% C.L.) in the three representative cases of ${\rm f}_{{\rm
Be}} =~$0.7; 1.0; 1.3:
$$\nu_{e} \leftrightarrow \nu_{\mu (\tau)}:~~~~~~~~~~~~~~~~~~
{\rm f}_{{\rm Be}} = 0.7,~~~0.35 \ltap {\rm f}_{{\rm B}} \ltap 3.4,
{}~~~~~~~~~~~~~~~~~~~~~~~~~~~~\eqno (13a)$$
$$~~~~~~~~~~~~~~~~~~~~~~~{\rm f}_{{\rm Be}} = 1.0,~~~ 0.35
\ltap {\rm f}_{{\rm
B}} \ltap 3.4~~~~~~~~~~~~~~~~~~~~~~~~~\eqno (13b)$$
$$~~~~~~~~~~~~~~~~~~~~~~~~~~~~~~~~~~~~~~~~~{\rm f}_{{\rm Be}} =
1.3,~~~ 0.35
\ltap {\rm f}_{{\rm B}} \ltap 0.43~~ {\rm and}~~ 0.46
\ltap {\rm f}_{{\rm B}} \ltap 3.4.~~~~~\eqno (13c)$$

\noindent
The allowed regions of values of $\Delta m^2$ and $\sin^22\theta$
corresponding to the solutions (13a), (13b) and (13c) are shown in
Figs.  1a, 1b and 1c, respectively.

    As Figs. 1a--1c illustrate, a deviation of the $^{8}$B and
$^{7}$Be neutrino fluxes from the values corresponding to ${\rm
f}_{{\rm B}} = {\rm f}_{{\rm Be}} = 1$ leads to two effects: i)
noticeable shift (and change in size) of the allowed $\Delta m^2 -
\sin^22\theta$ regions in the case ${\rm f}_{{\rm B}} = {\rm f}_{{\rm
Be}} = 1$ towards smaller (${\rm f}_{{\rm B}} <~1$) or larger (${\rm
f}_{{\rm B}} >~1$) values of $\sin^22\theta$ with the allowed values
of $\Delta m^2$ remaining practically within the interval of the ${\rm
f}_{{\rm B}} = {\rm f}_{{\rm Be}} = 1$ solution, $4.4\times
10^{-11}~{\rm eV}^2 \ltap \Delta m^2 \ltap 9.8\times 10^{-11}~{\rm
eV}^2$, and ii) appearance of new allowed regions of values of $\Delta
m^2$, i.e. of new solutions, at $\Delta m^2 \ltap 4.1\times
10^{-11}~{\rm eV}^2$. We find two such new solutions (see Figs. 1a --
1c):

\noindent (A) \hskip 0.2truecm for $0.7\leq {\rm f}_{{\rm Be}}
\leq 1.3$
and $0.35 \ltap {\rm f}_{{\rm B}} \ltap 0.44$ with $\Delta m^2$ and
$\sin^22\theta$ lying in the intervals $$4.7\times 10^{-12}~{\rm eV}^2
\ltap \Delta m^2 \ltap 6.5\times 10^{-12}~{\rm eV}^2,~~~ 0.71 \ltap
\sin^22\theta \leq 1.0,~~~~~~~~~~~\eqno(14)$$

\noindent (B) \hskip 0.2truecm for any value of
${\rm f}_{{\rm Be}}$ from the interval (9) (for ${\rm f}_{{\rm Be}} =
0.7$) and $0.45~(0.42) \ltap {\rm f}_{{\rm B}} \ltap 0.65~(0.66)$, and
for $\Delta m^2$ and $\sin^22\theta$ having values within the
intervals
\footnote{\tighten For ${\rm f}_{{\rm Be}} \cong 0.7$
there are also new solutions in the region $10^{-10}~{\rm eV}^2 <
\Delta m^2 < 10^{-9}~{\rm eV}^2$ and $0.62 \ltap \sin^22\theta \ltap
0.80$, representing three very narrow strips (almost lines) of allowed
values of $\Delta m^2$ and $\sin^22\theta$ in the $\Delta m^2 -
\sin^22\theta$ plane (see Fig. 1a).  However, these solutions are not
stable with respect to variations of ${\rm f}_{{\rm Be}}$ and
disappear when ${\rm f}_{{\rm Be}}$ is slightly increased (they do not
exist for ${\rm f}_{{\rm Be}} = 1.0$, for example). We shall not
discuss them further.}  $$3.2\times 10^{-11}~{\rm eV}^2 \ltap \Delta
m^2 \ltap 4.0\times 10^{-11}~{\rm eV}^2,~~~ 0.59 \ltap \sin^22\theta
\leq 1.0.~~~\eqno(15)$$
\noindent Both solutions (A) and (B) are stable with respect to changes
of ${\rm f}_{{\rm Be}}$
\footnote{\tighten The possibility of a "low" $^{8}$B neutrino flux
solution for ${\rm f}_{{\rm Be}} = 1.0$ at $\Delta m^2 = 6.0\times
10^{-12}~{\rm eV}^2$ and $\sin^22\theta = 0.8$ was suggested on the
basis of qualitative arguments in ref. [29]. Our results show that at
95\% C.L. the indicated point in the relevant parameter space is
marginally excluded by the current solar neutrino data
.}.
Nevertheless the regions of values of $\Delta m^2$ and
$\sin^22\theta$ of these solutions vary somewhat with ${\rm f}_{{\rm
Be}}$: eqs. (14) and (15) represent the largest intervals and
correspond practically to ${\rm f}_{{\rm Be}} \cong 0.7$.

   Let us discuss the above results. The probability that a solar
electron neutrino with energy E will not change into $\nu_{\mu
(\tau)}$ (or $\nu_s$) on its way to the Earth when
$\nu_e\leftrightarrow\nu_{\mu (\tau)}$ ($\nu_e\leftrightarrow\nu_s$)
oscillations take place, can be written in the form:
\vskip -0.4truecm
$${\rm P_{osc}(E; R(t))} = 1 - {1\over 2}
\sin^22\theta~\bigl[ 1 - \cos 2\pi {\rm R(t)\over
L_v}\bigr],\eqno(16)$$
\noindent where ${\rm L_v = 4\pi E}/\Delta m^2$ is the oscillation
length in vacuum, $${\rm R(t) = R_0~\bigl[ 1 - \epsilon\cos 2\pi
{t\over T}}\bigr],\eqno(17)$$
\noindent is the Sun--Earth distance at time t of the year
(T = 365 days), ${\rm R_0} = 1.4966\times 10^8$ km and $\epsilon =
0.0167$ being the mean Sun--Earth distance and the ellipticity of the
Earth orbit around the Sun.  The term with the $\epsilon$ factor in
eq. (17), as is well known [2-4,25], is a source of seasonal variation
effects in the case of the vacuum oscillation solution of the solar
neutrino problem
\footnote{\tighten Detailed predictions for the seasonal variation
effects in the present and future solar neutrino experiments in the
case of the ${\rm f}_{{\rm B}}\cong 1$, ${\rm f}_{{\rm Be}}\cong 1$
vacuum oscillation solution are given in ref. [25].}. Since we are
dealing in the analysis of interest with experimental results averaged
over at least few complete years of measurements, the relevant
probability is actually the probability (16) averaged over a period of
1 year, ${\rm
\bar{P}_{osc}(E; R_0,\epsilon)}$.  If $2\pi {\rm R_0/L_v} \ltap
2.5\pi$, ${\rm \bar{P}_{osc}(E; R_0,\epsilon)}$ practically coincides
with the probability (16) in which the parameter $\epsilon$ is
formally set to zero, i.e., with ${\rm P_{osc}(E; R_0)}$ (see Figs. 2a
and 2b). This implies that for the values of $\Delta m^2 \ltap
10^{-10}~{\rm eV}^2$ of interest one has ${\rm \bar{P}_{osc}(E;
R_0,\epsilon)} \cong {\rm P_{osc}(E; R_0)}$ for all neutrinos with
energy E$~\gtap~$3 MeV, i.e., for the dominant fraction of the $^{8}$B
neutrino flux. If, however, $2\pi {\rm R_0/L_v} >> 2.5\pi$, the effect
of the averaging can be quite dramatic for the oscillation's amplitude
\footnote{\tighten
It is not difficult to convince oneself that up to corrections which
do not exceed $5\times 10^{-3}$ the periods of
$\nu_e\leftrightarrow\nu_{\mu (\tau)}$ oscillations implied by the
probabilities ${\rm \bar{P}_{osc}(E; R_0,\epsilon)}$ and ${\rm
P_{osc}(E; R_0)}$ coincide.} and (for a given $\sin^22\theta$) ${\rm
\bar{P}_{osc}(E; R_0,\epsilon)}$ can differ considerably from ${\rm
P_{osc}(E; R_0)}$, as Figs. 2a and 2b illustrate.

     Consider first the solutions at $\Delta m^2 \gtap 4.4\times
10^{-11}~{\rm eV}^2$, i.e., in the region in which the ${\rm f}_{{\rm
B}} = {\rm f}_{{\rm Be}} = 1$ solution lies. The allowed regions
corresponding to these solutions converge continuously (changing their
shape and dimensions) to the allowed region in the case ${\rm f}_{{\rm
B}} = {\rm f}_{{\rm Be}} = 1$ when ${\rm f}_{{\rm B}}$ and ${\rm
f}_{{\rm Be}}$ are varied continuously from the values they have for a
given solution to 1. The new solutions (A) and (B) identified above
are "disconnected" from the ${\rm f}_{{\rm B}} = {\rm f}_{{\rm Be}} =
1$ solution and disappear when ${\rm f}_{{\rm B}}$ and ${\rm f}_{{\rm
Be}}$ change continuously from 0.35 to 1.

   {}For $\Delta m^2 \gtap 4.4\times 10^{-11}~{\rm eV}^2$ and for the
energies of the pp neutrinos, ${\rm E} \leq 0.42~$MeV, the cosine term
in the expression for the probability (16) is a fastly oscillating
function of E. Therefore the integration over the neutrino energy E in
the contribution of the pp neutrinos to the signal in the Ga--Ge
experiments suppresses the part of the contribution containing the
cosine term and one has effectively in this case [2,3] ${\rm
P^{pp}_{osc}(E; R_0)} \cong 1 - {1\over 2}\sin^22\theta$. This implies
that depending on the value of $\sin^22\theta$ the pp $\nu_e$ flux is
suppressed approximately by a factor (0.50 -- 0.70).  The values of
$\Delta m^2$ for which the minimal values of $\sin^22\theta$ allowed
by the data occur (see Figs. 1a -- 1c) are determined by the condition
$\cos 2\pi {\rm R_0/ L_v} = -1$ for E = 0.862 MeV -- the energy of the
dominant $^{7}$Be neutrino component of the solar neutrino flux; they
correspond for a given $\sin^22\theta$ to a maximal possible
suppression of the flux of the 0.862 MeV $^{7}$Be electron neutrinos
at the Earth surface due to the oscillations
$\nu_e\leftrightarrow\nu_{\mu (\tau)}$.  Actually, the suppression of
the 0.862 MeV $^{7}$Be $\nu_e$ flux due to the vacuum oscillations is
not very strong and can practically be absent in the case of the
solution under discussion, the relevant suppression factor ranging
from approximately 0.30 (possible when ${\rm f}_{{\rm Be}} = 1.3)$ to
0.98 (possible for ${\rm f}_{{\rm Be}} \cong (0.7 - 1.0)$ and ${\rm
f}_{{\rm B}} \cong 1)$.

  The maximal value of ${\rm f}_{{\rm B}}$, ${\rm max~f}_{{\rm B}}
\cong 3.4$, for which the $\nu_{e} \leftrightarrow \nu_{\mu (\tau)}$
oscillations provide a description of the solar neutrino data is
determined primarily by the Kamiokande result (2) (as its independence
on ${\rm f}_{{\rm Be}}$ indicates) and by the specific dependence of
${\rm P_{osc}(E; R_0)}$ (see eq. (16)), on the solar neutrino energy
E. It can be understood qualitatively by considering the constraints
the Kamiokande data imply in this particular case. For any fixed
$\Delta m^2 \ltap 10^{-10}~{\rm eV}^2$, the probability ${\rm
P_{osc}(E;R_0)}$ has at most one minimum in the interval of $^{8}$B
neutrino energies $7.5~{\rm MeV}\ltap {\rm E}\leq 14.4~{\rm MeV}$
relevant to the Kamiokande experiments (see Figs. 2a and 2b). The
suppression of the integral in the expression (6) for R(K) is maximal
when the minimum of ${\rm P_{osc}(E;R_0)}$ occurs at ${\rm E =
E_{min}}$, $7.5~{\rm MeV} < {\rm E_{min}} < 14.4~{\rm MeV}$. In this
case ${\rm P_{osc}(E;R_0)}$ increases monotonically with the change of
E in the indicated interval both for ${\rm E < E_{min}}$ and ${\rm E >
E_{min}}$. This implies that there exists a maximal possible
suppression of the flux of $^{8}$B electron neutrinos with ${\rm E}
\gtap 7.5~{\rm MeV}$ (i.e., of the integral in the right hand side of
eq. (6)) due to ${\rm P_{osc}(E; R_0)}$, and hence a maximal possible
value of ${\rm f}_{{\rm B}}$ for which the vacuum $\nu_{e}
\leftrightarrow \nu_{\mu (\tau)}$ oscillations can provide an
explanation of the Kamiokande result (2). The value one obtains
numerically, ${\rm max~f}_{{\rm B}} \cong 3.4$, is somewhat smaller
than the upper bound (8) and is reached for, e.g., ${\rm
f}_{{\rm B}} = 1$ at $\Delta m^2 \cong 8.2\times 10^{-11}~{\rm eV}^2$
and $\sin^22\theta = 1.0$.  For these values of $\Delta m^2$ and
$\sin^22\theta$ one has min${\rm P_{osc}(E_{min}; R_0)} = 0$ and ${\rm
E_{min}} \cong 9.5~{\rm MeV}$, and the suppression of the integral in
eq. (6)
corresponds effectively to a constant factor ${\rm [P_{osc}(E;
R_0)}~+~0.16(1 - {\rm P_{osc}(E; R_0))]}
\cong 0.2$. The value of $\Delta m^2$ for which the solution for
${\rm f}_{{\rm B}} \cong 3.4$ exists depends, although weakly, on the
value of ${\rm f}_{{\rm Be}}$ chosen within the interval (9) because
${\rm \bar{P}_{osc}(E; R_0,\epsilon)}$ for E$=$0.862 MeV is very
sensitive to even small changes of $\Delta m^2$ in the vicinity of
$8.2\times 10^{-11}~{\rm eV}^2$, as Fig. 2b (after the necessary
rescaling of the values of E on the horizontal axis by the ratio
($10^{-10}/(8.2\times 10^{-11}) \cong 1.2$) illustrates (we have:
$2\pi R_0/L_v = 37.9018 (\Delta m^2/10^{-10}{\rm eV}^2)(1{\rm
MeV}/{\rm E})$ and in this case, for instance, ${\rm
\bar{P}_{osc}(E=0.862~ MeV; R_0,
\epsilon)} = {\rm P_{osc}(E=0.862~ MeV; R_0)} \cong 0.50~(0.70)$
for $\Delta m^2 = 8.2~(8.3) \times 10^{-11}~{\rm eV}^2$ and
$\sin^22\theta = 1.0$). Therefore the necessary suppression of the
$^{7}$Be contribution to the signals in the Cl--Ar and the Ga--Ge
experiments for the different values of ${\rm f}_{{\rm Be}}$
considered is achieved by small changes of the value of $\Delta m^2$
around the value $8.2\times 10^{-11}~{\rm eV}^2$. These changes do not
affect the suppression of the contributions of the $^{8}$B neutrinos
in R(Ar) and R(K), required by the Cl--Ar and Kamiokande data.

   In a similar way one can understand the minimal values of ${\rm
f}_{{\rm B}}$, ${\rm min~f}_{{\rm B}} \cong 0.51;~0.54;~0.57$
corresponding to ${\rm f}_{{\rm Be}} = 0.7;~1.0;~1.3$, for which there
exists at 95\% C.L.  a $\nu_{e} \leftrightarrow \nu_{\mu (\tau)}$
oscillation solution in the region $\Delta m^2\gtap 4.4\times
10^{-11}~{\rm eV}^2$. Depending on ${\rm f}_{{\rm Be}}$ these
solutions occur for values of $\Delta m^2 \cong (4.3 - 5.1)\times
10^{-11}~{\rm eV}^2$ and for $\sin^22\theta \cong (0.57 - 0.75)$ (see
Figs. 1a -- 1c). The minimal values of ${\rm f}_{{\rm B}}$ allowed by
the data are again determined by the Kamiokande result, and by the
specific energy dependence of ${\rm P_{osc}(E; R_0)}$ for values of
$\Delta m^2$ in the vicinity of $\Delta m^2 \cong 4.9\times
10^{-11}~{\rm eV}^2$. Indeed, for $\Delta m^2 = (4.3 - 5.1)\times
10^{-11}~{\rm eV}^2$ and for the energies of $^{8}$B neutrinos
$7.5~{\rm MeV \ltap E \leq 14.4~MeV}$ detected by the Kamiokande
experiments, ${\rm P_{osc}(E;R_0)}$ is a monotonically (rather
steeply) increasing function of E and for $\sin^22\theta =
0.60~(0.75)$ one has $0.5~(0.4)\ltap {\rm P_{osc}(E;R_0)}\ltap
0.8~(0.7)$ (see Figs. 2a and 2b). In this case and, e.g., for ${\rm
f}_{{\rm Be}} = 0.7$, the maximal suppression due to ${\rm
P_{osc}(E;R_0)}$ of the integral in the expression for R(K), eq. (6),
corresponds effectively to a constant factor $[{\rm P_{osc}(E;R_0)} +
0.16(1 - {\rm P_{osc}(E;R_0)})] \cong 0.7$.  The Kamiokande data then
imply (95\% C.L.) ${\rm f}_{{\rm B}}\gtap 0.52~$.  The minimal value
of ${\rm f}_{{\rm B}}$ one obtains depends somewhat on the value of
${\rm f}_{{\rm Be}}$ because the requisite suppression of the $^{7}$Be
0.862 MeV electron neutrino flux (and of the contributions of the
0.862 MeV $^{7}$Be neutrinos to R(K) and R(Ge)) is achieved now by an
adjustment of the value of $\sin^22\theta$ within the interval 0.57 --
0.75 (rather than by changing $\Delta m^2$), which in turn leads to a
non-negligible change of ${\rm P_{osc}(E;R_0)}$ for $7.5~{\rm MeV
\ltap E \leq 14.4~MeV}$. In the case of the solution with ${\rm
f}_{{\rm Be}} = 0.7$ and ${\rm f}_{{\rm B}} = 0.53$, for instance, the
pp and 0.862 MeV $^{7}$Be electron neutrino fluxes are suppressed due
to the oscillations by the factors 0.70 and 0.46, respectively, while
the pep (CNO) neutrino flux (fluxes) is not (are mildly)
suppressed. The predictions for the signals in the Cl--Ar and Ga--Ge
detectors read in this case ${\rm R(Ar) \cong 2.7~SNU}$ and ${\rm
R(Ge) \cong 74~SNU}$.

  It should be noted that the allowed regions found at 95\% C.L. for
${\rm f}_{{\rm Be}} \cong (0.7 - 1.3)$ and ${\rm f}_{{\rm B}} \cong
(0.8 - 1.2)$ in the present study lie practically all within the
allowed regions one obtains at 95\% C.L.  in the reference model [7]
when the estimated uncertainties in the theoretical predictions are
included in the analysis.

  Let us add finally that the minimal values of the $\chi^2-$function
for the solution under discussion in the three cases ${\rm f}_{{\rm
Be}} = 0.7; 1.0; 1.3$ respectively read 0.67; 0.51; 0.53 (for 2 d.f.)
and take place at ${\rm f}_{{\rm B}} = 2.2; 2.4; 2.4$ for $\Delta m^2
= (7.2; 7.4; 7.5)\times 10^{-11}~{\rm eV^2}$ and $\sin^22\theta =
1$. For ${\rm f}_{{\rm B}} \leq 1$, ${\rm min}~\chi^2 = 2.9$
\footnote{\tighten In contrast, ${\rm min}~\chi^2 = 0.25$ in the
case of the MSW small mixing-angle $\nu_{e} \rightarrow \nu_{\mu
(\tau)}$ solution.}  and is reached for ${\rm f}_{{\rm B}} = 1$,
$\Delta m^2 = 6.1\times 10^{-11}~{\rm eV^2}$ and $\sin^22\theta =
0.86$. In the case of solution (A), eq.(14), and for ${\rm f}_{{\rm
Be}} = 0.7~(1.0)$, ${\rm min}~\chi^2 = 4.4~(4.5)$ and corresponds to
${\rm f}_{{\rm B}} = 0.39$, $\Delta m^2 = 5.4~(5.6)\times
10^{-12}~{\rm eV^2}$ and $\sin^22\theta = 1.0$. The ${\rm min}~\chi^2$
value is somewhat larger for solution (B): for ${\rm f}_{{\rm Be}} =
1.0$, for instance, one has ${\rm min}~\chi^2 = 5.3$.
\vskip 0.3cm
\leftline{
\bf 3.1.2 Physical Implications of the New Low $\Phi_{{\rm B}}$
   Solution}
\vskip 0.2cm

   The physical implications of the $\nu_{e} \leftrightarrow \nu_{\mu
(\tau)}$ oscillation solution for ${\rm f}_{{\rm Be}} \cong 1$ and
${\rm f}_{{\rm B}} \cong 1$ and values of $\Delta m^2$ in the interval
$4.4\times 10^{-11}~{\rm eV}^2 \ltap \Delta m^2 \ltap 10^{-10}~{\rm
eV}^2$ have been extensively discussed in the literature (see
refs. [2,25], [23] and the articles by A. Acker et al. and by
V. Barger et al. quoted in ref. [3]).  The solutions we have found in
the same $\Delta m^2$ region for ${\rm f}_{{\rm Be}} \neq 1$ and ${\rm
f}_{{\rm B}} \neq 1$ lead to generically similar implications and we
shall not consider them here.

   Of the two new solutions (A), eq. (14), and (B), eq. (15), solution
(A) is more interesting phenomenologically, has a lower
$\chi^2-$value, and therefore we shall discuss only it, although
rather briefly. For the values of $\Delta m^2$ from the interval given
in (14) one has: i) ${\rm \bar{P}_{osc}(E; R_0,
\epsilon)}\gtap 0.97~(0.94)$ for
${\rm E} \geq 7.5~(5.0)~{\rm MeV}$, ii) for neutrino energies in the
vicinity of 0.862 MeV ${\rm \bar{P}_{osc}(E; R_0,\epsilon)}$ has its
first local minimum when E decreases from values for which ${\rm
\bar{P}_{osc}(E; R_0,\epsilon)} \cong 1$ (see Figs. 2), and iii) the
first local maximum of ${\rm \bar{P}_{osc}(E; R_0,\epsilon)}$ as E
decreases below 0.862 MeV occurs in the interval $0.23~{\rm MeV} \ltap
{\rm E} \ltap 0.42~{\rm MeV}$
\footnote {\tighten One can explain the minimal (maximal) value of
${\rm f}_{{\rm B}}$ for which solution (A) exists, the reason for the
difference between the maximal allowed values of $\Delta m^2$ (and the
values themselfs) for a given ${\rm f}_{{\rm Be}}$ in the cases ${\rm
f}_{{\rm B}} = 0.35$ and ${\rm f}_{{\rm B}} = 0.38$ etc., in a similar
way we did it earlier, e.g., for the maximal value of ${\rm f}_{{\rm
B}}$ allowed by the data and the corresponding values of $\Delta
m^2$.}. Correspondingly, if solution (A) is realized, the signals due
to the $^{8}$B neutrinos in the present (and the future SNO [30] and
Super-Kamiokande [31]) detectors will not practically be affected by
the vacuum $\nu_{e} \leftrightarrow \nu_{\mu (\tau)}$ oscillations
(remember that for solution (A) ${\rm f}_{{\rm B}} \cong (0.35 -
0.44)$), the 0.862 MeV $^{7}$Be $\nu_e$ flux will be suppressed by the
suppression factor (0.06 - 0.26), while the signal due to the pp
electron neutrinos in the Ga--Ge detectors will be mildly reduced by a
suppression factor not smaller than 0.7. Thus, from the point of view
of how the different components of the solar neutrino flux are
affected, the vacuum oscillation solution (A) is very similar to the
low ${\rm f}_{{\rm B}}$ MSW $\nu_{e} \rightarrow \nu_{\mu (\tau)}~$
[21] or $\nu_{e} \rightarrow \nu_{s}~$ [26] transition nonadiabatic
solution. However, some of the physical implications of the two
solutions differ considerably. In particular, i) the predicted
distortion of the spectrum of the pp neutrinos is much stronger in the
case of the vacuum oscillation solution (A) \footnote{\tighten It is
also very different from the distortion of the pp neutrino spectrum in
the case of the solutions with $4.4\times 10^{-11}~{\rm eV}^2 \ltap
\Delta m^2
\ltap 10^{-10}~{\rm eV}^2$ (see the third article quoted in
ref. [2]).} (Fig. 3a) than for the corresponding MSW nonadiabatic
solution, and ii) if solution (A) is valid, the $^{7}$Be and pp
electron neutrino fluxes at the Earth surface will exhibit seasonal
variations which cannot take place in the case of the MSW solutions
(see ref. [32] and the first article quoted in ref. [24]).  In what
follows we shall discuss briefly the seasonal variation effects
predicted in the case of solution (A).

  {}For $\Delta m^2 \leq 6.5\times 10^{-12}~{\rm eV^2}$ and ${\rm E}
\geq 0.233~{\rm MeV}~(0.217~{\rm MeV})$ one has: $2\pi \epsilon {\rm
R_0/L_v} \leq 0.18~(0.19) << 1$. Thus, in the case of solution (A) the
probability ${\rm P_{osc}(E; R(t))}$, eq. (16), can be represented as
a power series in the small parameter $(2\pi \epsilon ({\rm
R_0/L_v})\cos 2\pi ({t/T}))$.  Neglecting all the terms smaller than
$10^{-3}$ in this series we obtain: $${\rm P_{osc}(E; R(t))} \cong
{\rm P_{osc}(E; R_0)} + {\rm P_{seas}(E; R_0,
\epsilon,t)},~~\eqno(18)$$
\noindent where the term
$${\rm P_{seas}(E; R_0,\epsilon,t)} = {1\over 2}\sin^22\theta~[2\pi
\epsilon {\rm {R_0\over L_v}}
\sin 2\pi {\rm {R_0\over L_v}}]~\cos 2\pi {\rm {t\over T}} -
{1\over 4}\sin^22\theta~\cos 2\pi {\rm {R_0\over L_v}}~[2\pi \epsilon
{\rm {R_0\over L_v}}\cos 2\pi {\rm {t\over T}}]^2~~~\eqno(19)$$
\noindent is responsible for the seasonal variation effects of
interest.  Obviously, for fixed values of the parameters $\Delta m^2$
and $\sin^2 2\theta$, the difference between the values of ${\rm
P_{seas}(E; R_0,\epsilon, t)}$ in December/January (${\rm t} \cong 0$)
and June/July (${\rm t} \cong 0.5{\rm T}$) is the largest.

  {}For $^{8}$B neutrinos with ${\rm E}\geq 5~{\rm MeV~(6.44~MeV)}$ we
have $2\pi \epsilon {\rm R_0/L_v} \leq 8.2\times 10^{-3}~(6.3\times
10^{-3})$ and it follows from eqs. (18) and (19) that the seasonal
variation effect in the signal of the Super--Kamiokande (SNO) detector
will be too small to be observable. The effect can be much larger for
the signals due to the $^{7}$Be and/or pp neutrinos in the Ga--Ge,
BOREXINO [33] and HELLAZ [34] detectors.

  In the case of solution (A) one has for the predicted average rate
of Ge production per year in the Ga--Ge experiments for ${\rm f}_{{\rm
Be}} = 0.7; 1.0; 1.3$: ${\rm \bar{R}(Ge)
\ltap 80; 84; 88~SNU}$.
The difference between the rates of Ge production in December/January
(${\rm t} \cong 0$) and June/July (${\rm t \cong 0.5T}$), $\Delta {\rm
R_{seas}(Ge)}$, due to i) the term (19) in the vacuum oscillation
probability (18), and ii) the change of the neutrino fluxes with the
change of the Sun--Earth distance due to the standard geometrical
effect, as can be shown, satisfies: $4.0~{\rm SNU}\ltap \Delta {\rm
R_{seas} (Ge)} \ltap 8.7~(8.1)~{\rm SNU}$, the maximal value
corresponding to ${\rm f}_{{\rm Be}} = 1.3~(0.7)$. A convenient
relative measure of the predicted seasonal effect is the seasonal
(December/January -- June/July) asymmetry: $${\rm A_{seas}(Ge)} = {\rm
{R_0^2\over \bar{R}(Ge)} [ {R(Ge;t)\over R^2(t)}|_{t\cong 0} -
{R(Ge;t)\over R^2(t)}|_{t\cong 0.5T}]},~~\eqno (20)$$
\noindent where ${\rm R(Ge;t)~[R_{0}/R(t)]^2}$ is the rate of Ge
production at time t of the year and R(t) is given by eq. (17). For
solution (A) we have: $0.072 \ltap {\rm A_{seas}(Ge)} \ltap 0.13$, the
contribution due purely to the geometrical factor ${\rm R^{-2}(t)}$
being $4\epsilon = 0.0668$. For given $\Delta m^2$ and $\sin^22\theta$
the change of ${\rm A_{seas}(Ge)}$ with the change of ${\rm f}_{{\rm
Be}}$ is negligibly small.

  The corresponding seasonal (December/January -- June/July) asymmetry
in the signal due to the $^{7}$Be (0.862 MeV) neutrinos in the
BOREXINO detector is given (up to corrections smaller than $10^{-3}$)
by $${\rm A^{a}_{seas}(BOR)} = 0.0668 + {{0.79~2\pi \epsilon {\rm
R_0/L_v}}\over {0.21 + 0.79 {\rm P_{osc}(E; R_0)}}}~
\sin^22\theta~\sin 2\pi {\rm {R_0\over L_v}},~~\eqno (21)$$
\noindent where 0.0668 is the asymmetry in the absence of oscillations,
and we have used the fact that $\sigma_{\nu_{\mu
(\tau)}e}$(E)/$\sigma_{\nu_{e}e}$(E)~$ \cong 0.21$ for E = 0.862
MeV. Note that the asymmetry ${\rm A^{a}_{seas}(BOR)}$ does not depend
on the total flux of 0.862 MeV $^{7}$Be neutrinos.  Note also that in
the case of solution (A) for E = 0.862 MeV we have $\sin 2\pi {\rm
R_0/ L_v} > 0$ and the second term in eq. (21) is always positive. As
can be shown, the asymmetry ${\rm A^{a}_{seas}(BOR)}$ changes very
little with the variation of $\Delta m^2$ and $\sin^22\theta$ within
the allowed regions of values for the solution (A) and ${\rm
A^{a}_{seas}(BOR)}
\cong (0.11 - 0.13)$.

      The HELLAZ experiment [34] is envisaged to detect pp neutrinos
having energy ${\rm E \geq 0.217~MeV}$ and to measure their
spectrum. The experiment will be based on the $\nu - e^{-}$ elastic
scattering reaction. Since the energy of the incident pp neutrino in
each event will be reconstructed, one can define a seasonal asymmetry
in the signal of HELLAZ, generated by neutrinos having energy within a
given interval ${\rm E_1 \leq E \leq E_2}$ (${\rm E_1 \geq
0.217~MeV}$, ${\rm E_2 \leq 0.42~MeV}$): ${\rm A_{seas}(H; E_1,
E_2)}$. The expression for ${\rm A_{seas}(H; E_1, E_2)}$ can be
obtained formally from eq.  (20) by replacing ${\rm R(Ge;t)}$ and
${\rm \bar{R}(Ge)}$ with the corresponding quantities -- event rate at
time t of the year, ${\rm R(H; E_1, E_2, t)}$, and mean event rate per
year, ${\rm \bar{R}(H; E_1, E_2)}$, for HELLAZ.

    {}For solution (A) of interest and ${\rm E \geq 0.217~MeV}$ one
has $2\pi \epsilon {\rm R_{0}/L_{v}} \ltap 0.19$, and the seasonal
asymmetry in the total (neutrino energy integrated) signal of the
HELLAZ detector, ${\rm A^{a}_{seas}(H)}$, as numerical calculations
show, satisfies $0.018 \ltap {\rm A^{a}_{seas}(H)} \ltap 0.067$, the
smallest (the largest) value being reached for $\Delta m^2 \cong
5\times 10^{-12}~{\rm eV^2}$ ($6\times 10^{-12}~{\rm eV^2}$). Note
that for $\Delta m^2 \cong 5\times 10^{-12}~{\rm eV^2}$ the asymmetry
due to the $\nu_{e} \leftrightarrow \nu_{\mu (\tau)}$ oscillations
compensates to a large extent the geometrical one, rendering the net
asymmetry hardly observable. The absence of any seasonal variations in
the signal of SNO, or Super--Kamiokande, or BOREXINO, or HELLAZ
detector (constant in time event rate) is known [2,25] to be one of
the distinctive signatures of the vacuum oscillation solutions of the
solar neutrino problem.

     The comparatively small values of the asymmetry ${\rm
A^{a}_{seas}(H)}$ are a consequence of two circumstances. First, the
contribution of the $\nu_{\mu (\tau)}$ neutrinos (present in the pp
($\nu_e$) flux as a result of the oscillations) reduces the asymmetry
of interest. For a given energy E the expression for the latter
contains the factor $(1 - \sigma_{\nu_{\mu (\tau)}e}$(E)/
$\sigma_{\nu_{e}e}$(E)) which changes from 0.58 to 0.72 when E
increases from 0.217 MeV to 0.42 MeV \footnote{\tighten The same term
gives rise to the factor 0.79 in the expression for ${\rm
A^{a}_{seas}(BOR)}$, eq. (21).}. More importantly, for values of
$\Delta m^2$ from the interval (14), $\sin 2\pi {\rm R_0/ L_v}$
changes sign passing through zero in the interval $0.28~{\rm MeV \leq
E \leq 0.39~MeV}~$ when E varies from 0.217 MeV to 0.42 MeV
\footnote{\tighten For $5.4\times 10^{-12}~{\rm eV^2}\ltap \Delta m^2
\ltap 6.5\times 10^{-12}~{\rm eV^2}$, $\sin 2\pi {\rm R_0/ L_v}$
changes sign two times in the interval (0.217 MeV -- 0.42 MeV),
passing through a second zero located at $0.217~{\rm MeV \leq E \leq
0.26~MeV}~$. However, the effect of the presence of this second zero
of $\sin 2\pi {\rm R_0/ L_v}$ on ${\rm A^{a}_{seas}(H)}$ is less
important than the effect of the first zero located at ${\rm E \geq
0.28~MeV}$.}. As a result the two contributions to the asymmetry in
the energy integrated event rate, generated by the oscillations of pp
neutrinos having energy in the two intervals $0.217~{\rm MeV \leq E
\leq E_{0}}$ and ${\rm E_{0} \leq E \leq 0.42~ MeV}$, ${\rm E_{0} \geq
0.28~MeV}$ being the energy at which $\sin 2\pi {\rm R_0/ L_v} = 0$
\footnote{\tighten
Obviously, the value of ${\rm E_{0}}$ depends on the value chosen of
$\Delta m^2$.}, have opposite signs and compensate partially or
completely each other. A complete cancelation between the indicated
two contributions takes place, for instance, for $\Delta m^2 \cong
6\times 10^{-12}~{\rm eV^2}$, for which ${\rm E_{0} \cong 0.36~MeV}$.

   It should be evident from the above discussion that for solution
(A) the asymmetry ${\rm A^{a}_{seas}(H; 0.217~MeV, E_{0})}
\equiv {\rm A^{a}_{seas}(H; E \leq E_{0})}$  or
${\rm A^{a}_{seas}(H; E_{0}, 0.42~MeV)} \equiv {\rm A^{a}_{seas}(H; E
\geq E_{0})}$ can be larger than ${\rm A^{a}_{seas}(H)}$. Indeed, it
can be easily shown that either $|{\rm A^{a}_{seas}(H; E \leq E_{0})}|
> {\rm A^{a}_{seas}(H)}$, or $|{\rm A^{a}_{seas}(H; E \geq E_{0})}| >
{\rm A^{a}_{seas}(H)}$. One can have also $|{\rm A^{a}_{seas}(H; E
\leq E_{0})} - {\rm A^{a}_{seas}(H; E \geq E_{0})}| > {\rm
A^{a}_{seas}(H)}$. Note that the difference $[{\rm A^{a}_{seas}(H; E
\leq E_{0})} - {\rm A^{a}_{seas}(H; E \geq E_{0})}]$ is free from the
geometrical term and can be nonzero only if the pp neutrinos take part
in vacuum oscillations. Let us illustrate the above remarks with two
examples. For $\Delta m^2 = 6\times 10^{-12}~{\rm eV^2}$ (${\rm E_{0}
= 0.36~MeV}$) and $\sin^22\theta = 1$ one has: ${\rm A^{a}_{seas} (H)}
\cong 0.0668$ and ${\rm A^{a}_{seas}(H; E \leq 0.36~MeV)} \cong
0.11$. If $\Delta m^2 = 5\times 10^{-12}~{\rm eV^2}$, then ${\rm E_{0}
= 0.30~MeV}$ and in this case ${\rm A^{a}_{seas}(H)} \cong 0.018$,
${\rm A^{a}_{seas}(H; E \leq 0.30~MeV)} \cong 0.10$, ${\rm
A^{a}_{seas}(H; E \geq 0.30~MeV)} \ltap 10^{-3}$, and $[{\rm
A^{a}_{seas}(H; E \leq 0.30~MeV)} - {\rm A^{a}_{seas}(H; E \geq
0.30~MeV)}] \cong 0.10$.

   In the case of the solution with $4.4\times 10^{-11}~{\rm eV}^2
\ltap \Delta m^2 \ltap 10^{-10}~{\rm eV}^2$ the seasonal asymmetry in
the signals of the Ga--Ge, Super--Kamiokande (SNO) and BOREXINO
detectors due purely to the $\nu_{e} \leftrightarrow \nu_{\mu (\tau)}$
oscillations can be as large as 30\%, 14\% and 80\%, respectively, and
is predicted to be negligible for the energy integrated signal of the
HELLAZ detector [2,25].
\vskip 0.3cm
\leftline{\bf 3.2 Oscillations into Sterile Neutrino
$\nu_{e} \leftrightarrow \nu_{s}$}
\vskip 0.2cm
  The solar neutrino oscillations into sterile neutrino, $\nu_{e}
\leftrightarrow \nu_{s}$, were excluded in the case ${\rm f}_{{\rm
B}}\cong 1$ and ${\rm f}_{{\rm Be}}\cong 1$ at 99\% C.L.  as a
possible solution of the solar neutrino problem by the (mean event
rate) solar neutrino data which were available by March 1994
[4]. Since then updated results have been published by all operating
solar neutrino experiments. The current status of the hypothesis of
solar $\nu_{e} \leftrightarrow \nu_{s}$ oscillations, including the
cases ${\rm f}_{{\rm B}}\neq 1$ and ${\rm f}_{{\rm Be}}\neq1$, $0.7
\leq {\rm f}_{{\rm Be}}\leq 1.3$, is summarized graphically in
Figs. 4a - 4c. At 95\% C.L. and for $0.7 \leq {\rm f}_{{\rm Be}}\leq
1.3$ (${\rm f}_{{\rm Be}} = 0.7$) this possibility is not excluded by
the current solar neutrino data only for $$0.35 \ltap{\rm f}_{{\rm
B}}\ltap 0.43~(0.44)~~~\eqno(22)$$
\noindent and values of $\Delta m^2$ and
$\sin^22\theta$ in the intervals $$4.8\times 10^{-12}~{\rm eV}^2 \ltap
\Delta m^2 \ltap 6.2\times 10^{-12}~{\rm eV}^2,~~~ 0.74 \ltap
\sin^22\theta \leq 1.0.~~~~~~~~~~~\eqno(23)$$
\noindent This solution is stable with respect to variations of
${\rm f}_{{\rm Be}}$ within the interval (9). Obviously, it is a
$\nu_{e}
\leftrightarrow \nu_{s}$ oscillation analog of the $\nu_{e}
\leftrightarrow
\nu_{\mu (\tau)}$ oscillation solution (A)
(compare eqs. (22), (23) with eq. (14)). At 95\% C.L. and for ${\rm
f}_{{\rm Be}} \cong 0.7$ there exist also two other allowed regions at
larger values of $\Delta m^2$ for ${\rm f}_{{\rm B}} \cong 0.42$ and
${\rm f}_{{\rm B}} \cong 0.50$ (see Fig. 4a), but they are very small
and disappear when ${\rm f}_{{\rm Be}} > 0.7$. If one increases the
required C.L. of the description of the data to 98\%, solution exists
for $0.32
\ltap{\rm f}_{{\rm B}}\ltap 1.8$ if ${\rm f}_{{\rm Be}} =
(0.7 - 1.3)$. The corresponding new allowed regions are scattered over
the area $4.0\times 10^{-12}~{\rm eV}^2 \ltap \Delta m^2 \ltap
2.0\times 10^{-10}~ {\rm eV}^2$, $0.50\ltap \sin^22\theta \leq
1.0$. These regions diminish in size considerably or completely
disappear as ${\rm f}_{{\rm Be}}$ changes from 0.7 to 1.3: for ${\rm
f}_{{\rm Be}} = 1.3$ most of the remaining new regions are in the form
of narrow strips (see Fig. 4c).

   Let us consider briefly the physical implications of the solution
(22) -- (23) for the future solar neutrino experiments. The
deformation of the pp neutrino spectrum (Fig. 3b) is quite strong and
differs somewhat from the deformation in the case of the $\nu_{e}
\leftrightarrow
\nu_{\mu (\tau)}$ oscillation solution (A) (compare Figs. 3a and 3b).
In what regards the seasonal variation effects in the signals of the
Super--Kamiokande, SNO and the Ga--Ge experiments, they coincide with
those for the $\nu_{e} \leftrightarrow \nu_{\mu (\tau)}$ oscillation
solution (A), considered in Section 3.1.2. The predicted seasonal
variation effect in the signal due to the $^{7}$Be neutrinos in the
BOREXINO detector, however, is larger than in the corresponding case
of $\nu_{e} \leftrightarrow \nu_{\mu (\tau)}$ oscillations. The
seasonal (December/January -- June/July) asymmetry for the BOREXINO
detector is given now by the expression $${\rm A^{s}_{seas}(BOR)} =
0.0668 + {{2\pi \epsilon {\rm R_0/L_v}}\over {{\rm P_{osc}(E;
R_0)}}}~\sin^22\theta~\sin 2\pi {\rm {R_0\over L_v}},~~\eqno (24)$$
\noindent and can be as large as 42\%: one has
$0.18\ltap {\rm A^{s}_{seas}(BOR)}\ltap 0.42$. We find also that ${\rm
A^{s}_{seas}(BOR)} \gtap 1.5{\rm A^{a}_{seas}(BOR)}$. The difference
between the values of ${\rm A^{s}_{seas}(BOR)}$ and ${\rm
A^{a}_{seas}(BOR)}$, in particular, can be used to distinguish between
the $\nu_{e} \leftrightarrow
\nu_{\mu (\tau)}$ oscillation solution (A) and its
$\nu_{e} \leftrightarrow \nu_{s}$ oscillation analog in a solar model
independent way.

       The seasonal asymmetries ${\rm A^{s}_{seas}(H; E \leq E_{0})}$
and ${\rm A^{s}_{seas}(H; E \geq E_{0})}$ in the signal of the HELLAZ
detector tend also to be larger than the corresponding asymmetries in
the case of the $\nu_{e} \leftrightarrow \nu_{\mu (\tau)}$ oscillation
solution (A). For $\Delta m^2 = 5.0\times 10^{-12}~{\rm eV}^2~$ (${\rm
E_{0} = 0.30~MeV}$) and $\sin^22\theta = 1.0$, for instance, we have:
${\rm A^{s}_{seas}(H)} \cong -0.9\times 10^{-2}$, ${\rm
A^{s}_{seas}(H; E \leq 0.30~MeV)} \cong 0.12$, ${\rm A^{s}_{seas}(H; E
\geq 0.30~MeV)} \cong -0.04$, and $[{\rm A^{s}_{seas}(H; E \leq
0.30~MeV)} - {\rm A^{s}_{seas}(H; E \geq 0.30~MeV)}] \cong 0.16$. In
the case of $\Delta m^2 = 6.0\times 10^{-12}~{\rm eV}^2~$ (${\rm E_{0}
= 0.36~MeV}$) one obtains: ${\rm A^{s}_{seas}(H)} \cong 0.0668$, and
${\rm A^{s}_{seas}(H; E \leq 0.36~MeV)} \cong 0.15$.

\vskip 0.2cm
\leftline{\bf 4. ENERGY INDEPENDENT SUPPRESSION OF THE SOLAR}
\hskip 0.3cm {\bf NEUTRINO FLUX}
\vskip 0.3cm
    The possibility of universal (energy independent) suppression of
the pp, $^{7}$Be, pep, $^{8}$B and CNO neutrino fluxes can be realized
if solar neutrinos take part in $\nu_{e} \leftrightarrow \nu_{\mu
(\tau)}$ or $\nu_{e} \leftrightarrow \nu_{s}$ oscillations
characterized by $\Delta m^2 >> 10^{-4}~{\rm eV}^2$. The solar matter
effects for $\Delta m^2 >> 10^{-4}~{\rm eV}^2$ are negligible and
neutrinos propagate in the Sun as in vacuum.  The averaging over the
region of neutrino production, etc. in the indicated case renders the
oscillating term in the expression for the oscillation probability,
eq. (16), negligible and one effectively has ${\rm P_{osc}} = 1 -
1/2~\sin^22\theta$ for all components of the solar neutrino flux.  The
Voloshin, Vysotsky, Okun [35] solar $\nu_e$ spin precession scenario
also leads to the indicated type of reduction of the solar $\nu_e$
flux.

   In general one has to consider two possibilities: transitions (or
oscillations) into active neutrino, $\nu_{e} \rightarrow \nu_{\mu
(\tau)}$ or $\nu_{e} \rightarrow \bar{\nu}_{\mu (\tau)}$, and into
sterile neutrino $\nu_{e} \rightarrow \nu_{s}$. From the point of view
of the analysis of the solar neutrino data currently available, there
is no difference between the cases of $\nu_{e} \rightarrow \nu_{\mu
(\tau)}$ and $\nu_{e} \rightarrow \bar{\nu}_{\mu (\tau)}$ transitions
(or oscillations).  This follows from the fact that for ${\rm E \gtap
7.5~MeV}$ the cross--sections $\sigma_{\nu_{\mu (\tau)}e}{\rm (E)}$
and $\sigma_{\bar{\nu}_{\mu (\tau)}e}{\rm (E)}$ practically coincide.

   We have investigated the possibility that the solar neutrino
deficit is due to a suppression of the different components of the
solar neutrino flux by one and the same energy independent factor
R resulting
from $\nu_{e} \rightarrow \nu_{\mu (\tau)}$ ($\nu_{e} \leftrightarrow
\nu_{\mu (\tau)}$) or $\nu_{e} \rightarrow \bar{\nu}_{\mu (\tau)}$, or
from $\nu_{e} \rightarrow \nu_{s}$ ($\nu_{e} \leftrightarrow \nu_{s}$)
transitions (oscillations). There are two parameters in the
corresponding $\chi^2~$--analysis: R and ${\rm f}_{{\rm B}}$. They
were varied within the intervals: (0.0 -- 1.0) and (0.0 -- 5.0),
respectively.  The parameter ${\rm f}_{{\rm Be}}$ was assumed to have
a fixed value within the interval (9).

    Our analysis showed that for ${\rm f}_{{\rm Be}} = 0.7; 1.0; 1.3$
a neutrino energy independent suppression of the solar neutrino flux
resulting from $\nu_{e} \rightarrow \nu_{\mu (\tau)}$ ($\nu_{e}
\leftrightarrow \nu_{\mu (\tau)}$) or $\nu_{e} \rightarrow
\bar{\nu}_{\mu (\tau)}$ transitions (oscillations) is excluded by the
current solar neutrino data at 97\%; 98\%; 98\% C.L. The regions in
the R -- ${\rm f}_{{\rm B}}$ plane allowed in this case at 99\%
C.L. ($\chi^2 \leq 9.21$) are shown in Figs. 5a -- 5c. Finally, for
the indicated values of ${\rm f}_{{\rm Be}}$ the solar neutrino data
rule out the hypothesis of constant suppression of the solar neutrino
flux due to $\nu_{e} \rightarrow \nu_{s}$ ($\nu_{e} \leftrightarrow
\nu_{s}$) transitions (oscillations) at 99.0\%; 99.5\%; 99.7\% C.L.
\vskip 0.2cm
\leftline{\bf 5. CONCLUSIONS}
\vskip 0.3cm

      We have shown that the $\nu_{e} \leftrightarrow \nu_{\mu
(\tau)}$ vacuum oscillation solution of the solar neutrino problem is
stable with respect to changes in the predictions for the fluxes of
$^{8}$B and $^{7}$Be neutrinos. For low values of $\Phi_{{\rm B}}$
(${\rm f_{B} \cong 0.35 - 0.43}$) new $\nu_{e}
\leftrightarrow \nu_{\mu (\tau)}$ and $\nu_{e} \leftrightarrow
\nu_{s}$ oscillation solutions exist. We have discussed the physical
implications of these new solutions for the future solar neutrino
experiments. A second new $\nu_{e} \leftrightarrow \nu_{\mu (\tau)}$
oscillation solution has been found for values of ${\rm f_{B}}$
which lie within the interval ${\rm f_{B} \cong 0.45 - 0.65}$.
The current solar neutrino data exclude at 99 \% C.L.
the possibility of universal (energy independent) suppression of the
different components of the solar neutrino flux, caused by $\nu_{e}
\leftrightarrow \nu_{s}$ oscillations or $\nu_{e} \rightarrow \nu_{s}$
transitions. A similar suppression resulting from solar $\nu_e$
oscillations or transitions into an active neutrino ($\nu_{e}
\leftrightarrow \nu_{\mu (\tau)}$, $\nu_{e} \rightarrow \bar{\nu}_{\mu
(\tau)}$) is strongly disfavoured by the data: depending on the value
of ${\rm f_{Be}}$ it is excluded at 97\%--98\% C.L.

\vskip 0.3cm
\leftline{\bf Acknowledgements.} S.T.P. wishes to thank O. P\`ene and
the other colleagues from L.P.T.H.E., Universit\'e de Paris -- Sud,
where part of the work for this study has been done, for the kind
hospitality extended to him during his visit. P.I.K. thanks the
organizing committee of the Santa Fe workshop on ``Massive Neutrinos
and Their Implications'' during which part of this work was
completed. The work of S.T.P. was supported in part by the EEC grant
ERBCHRX CT930132. The work of P.I.K. was supported by a grant from the
Institute for Advanced Study.

\vskip 0.4cm
\centerline{\bf Figure Captions}
\medskip
\noindent
{\bf Figs. 1a -- 1c.} Regions of values of the parameters $\Delta m^2$
and $\sin^22\theta$ for which the solar neutrino data can be described
at 95\% C.L. in terms of $\nu_{e} \leftrightarrow \nu_{\mu (\tau)}$
oscillations of the solar $\nu_e$ for values of ${\rm f_{Be}} =
0.7~({\rm a}), 1.0~({\rm b}), 1.3~({\rm c})$, and for values of ${\rm
f_{B}}$ from the interval (0.35 -- 2.5).

\medskip

\noindent
{\bf Figs. 2a -- 2b.} The vacuum oscillation probability for the mean
distance between the Sun and the Earth, ${\rm P_{osc}(E; R_0)}~$ (a),
and the probability (16) averaged over a period of 1 year, ${\rm
\bar{P}_{osc}(E; R_0, \epsilon)}~$ (b), as function of the neutrino
energy E for $\Delta m^2 = 10^{-10}~{\rm eV^2}~$ and $\sin^22\theta =
0.8$.
\medskip

\noindent
{\bf Figs. 3a -- 3b.} The deformation of the normalized to one
spectrum of pp neutrinos in the cases of a) $\nu_{e} \leftrightarrow
\nu_{\mu (\tau)}$ and b) $\nu_{e} \leftrightarrow \nu_{s}$ oscillation
solutions (14) and (23), respectively. The dotted, dashed,
long--dashed, dash--dotted and long--dash--dotted lines correspond to
$\Delta m^2 = (5.2; 5.4; 5.6; 5.8; 6.0)\times 10^{-12}~{\rm eV^2}$ and
$\sin^22\theta = 1$.

\medskip
\noindent
{\bf Figs. 4a -- 4c.} The regions of values of the parameters $\Delta
m^2$ and $\sin^22\theta$ for which the solar neutrino data can be
described at 95\% C.L. (dashed lines) and 98\% C.L.  (solid lines) in
terms of $\nu_{e} \leftrightarrow \nu_{s}$ oscillations of the solar
$\nu_e$ for ${\rm f_{Be} = 0.7~({\rm a); 1.0~(b); 1.3~(c)}}$, and for
values of ${\rm f_{B}}$ from the interval (0.35 - 1.5).
\medskip

\noindent
{\bf Figs. 5a -- 5c.} The regions of values of the parameters R and
${\rm f_{B}}$ allowed at 99\% C.L. ({\bf $\chi^2 \le 9.21$}) by the
solar neutrino data in the case of universal (energy independent)
suppression of the different components of the solar neutrino flux by
one and the same factor R, caused by vacuum oscillations or
transitions of the solar neutrinos into an active neutrino ($\nu_{\mu
(\tau)}$ or $\bar{\nu}_{\mu (\tau)}$) in the three cases ${\rm f_{Be}
= 0.7~(a); 1.0~(b); 1.3~(c)}$.

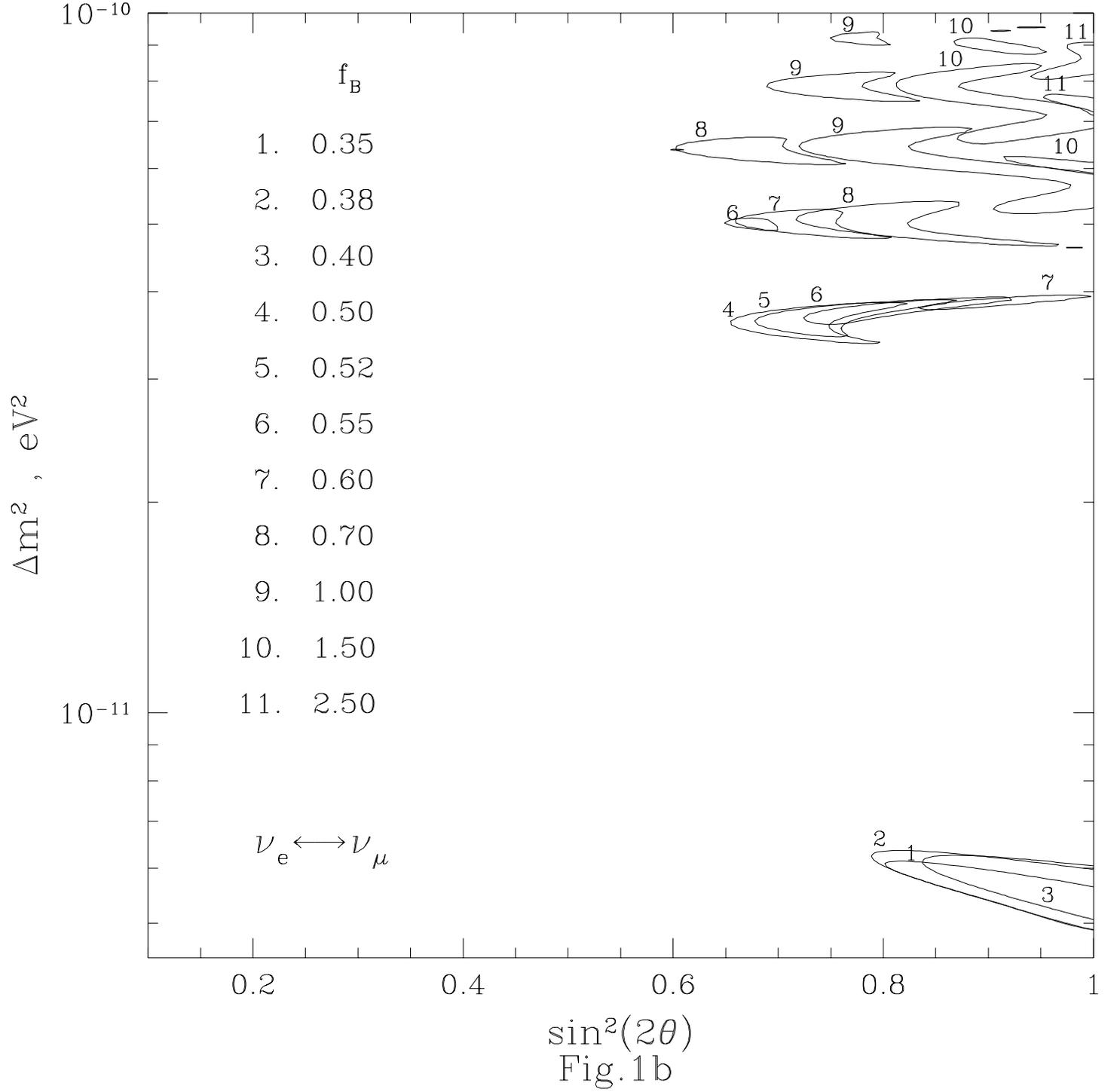
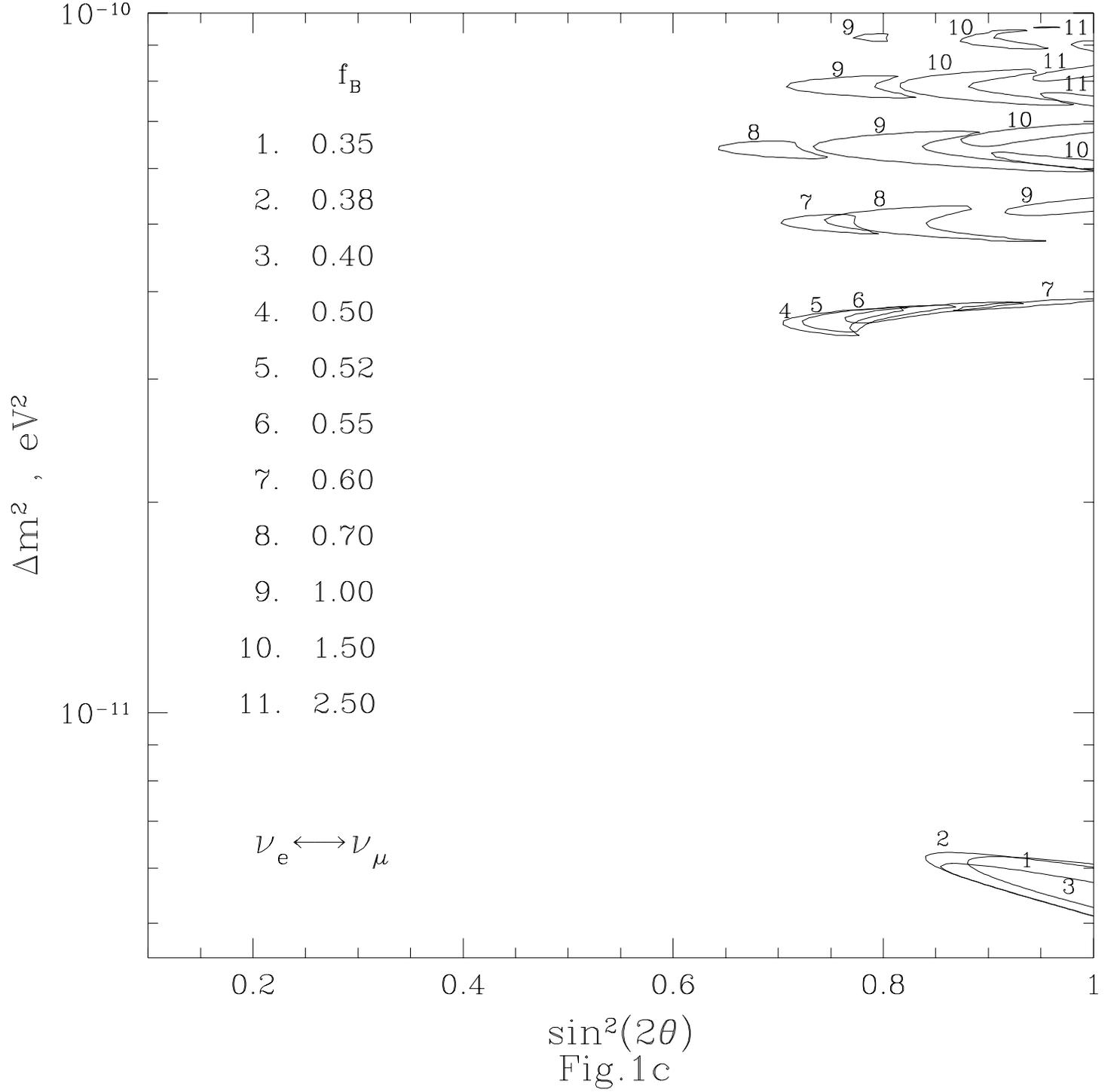
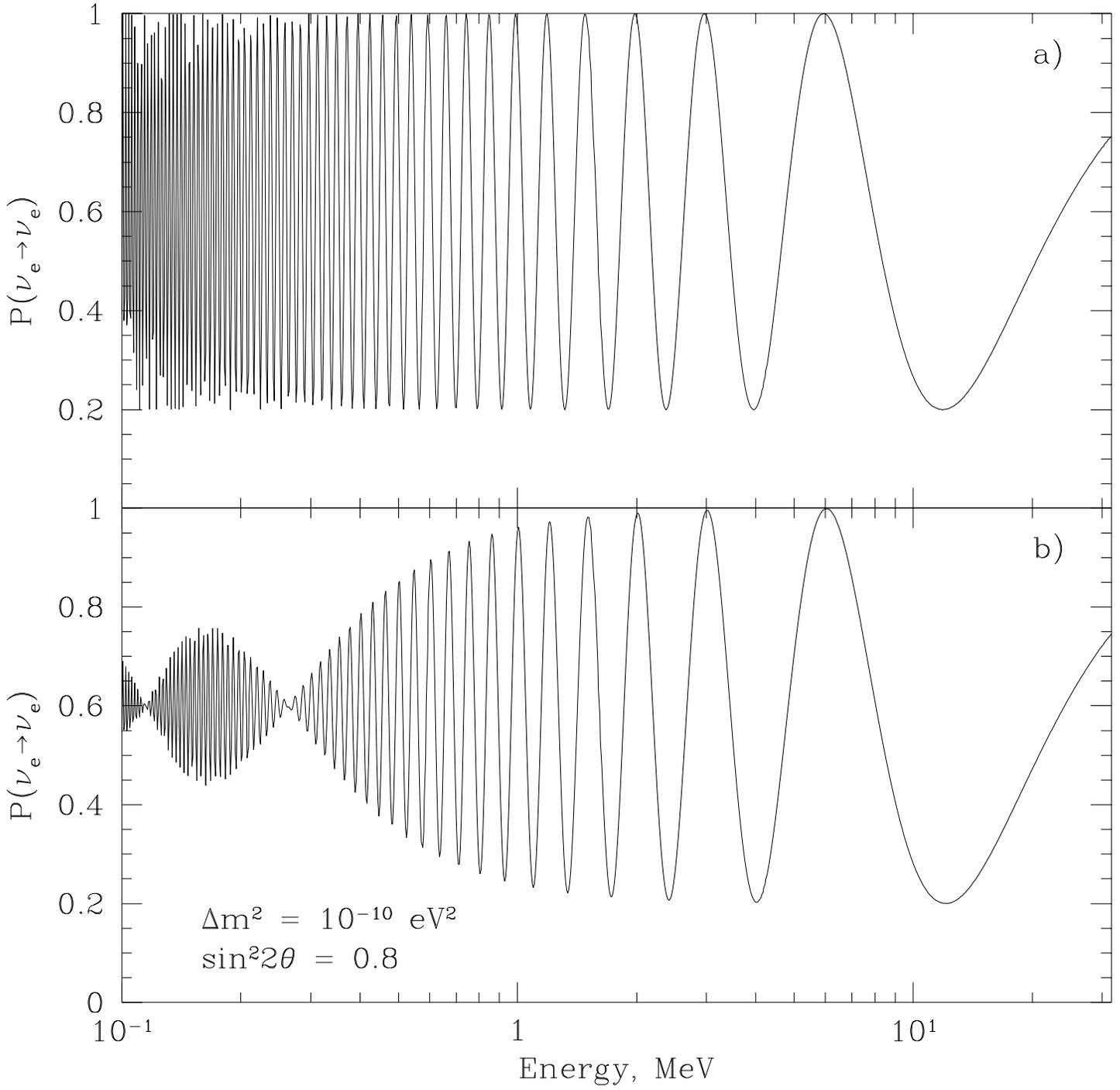
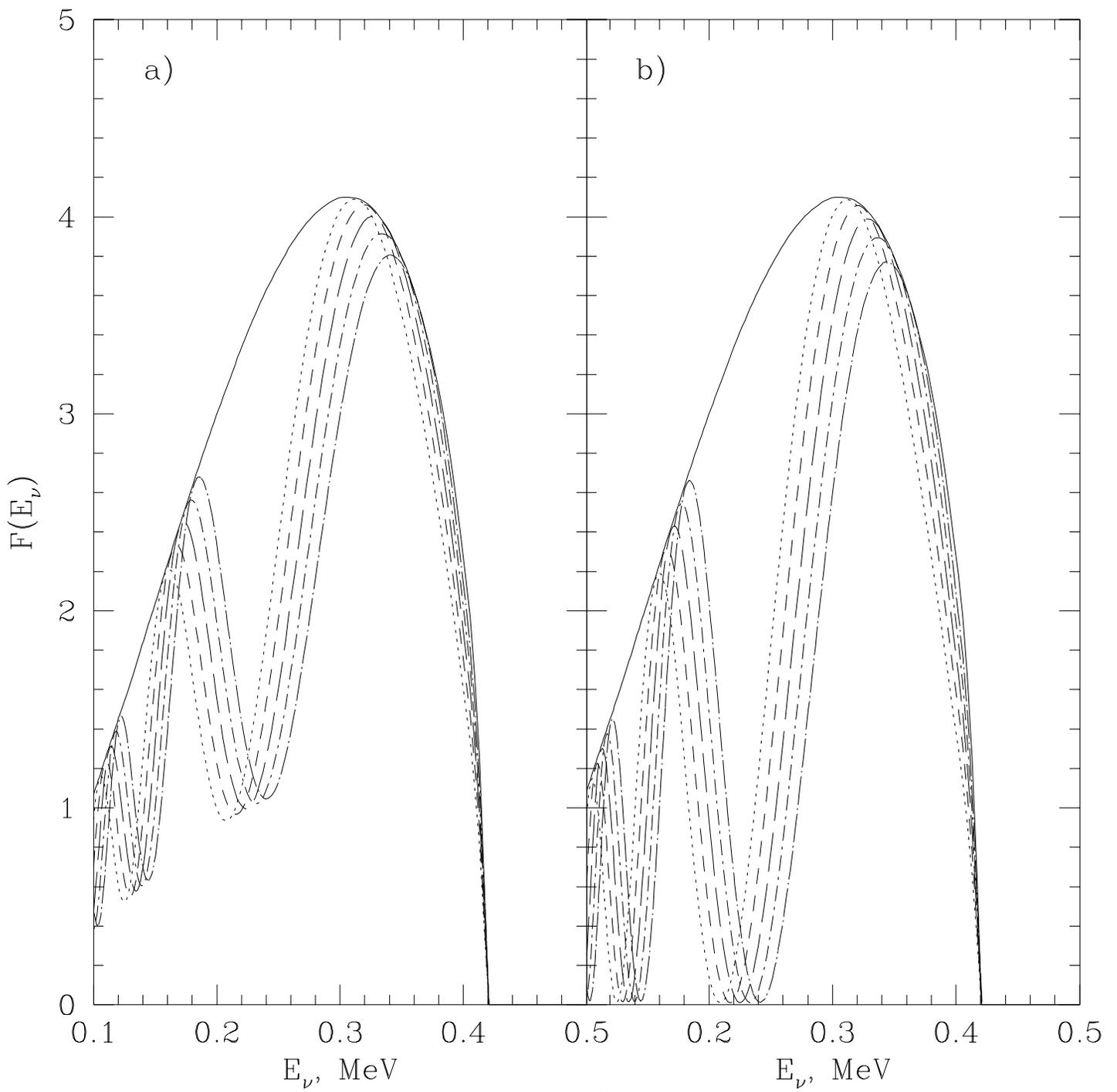
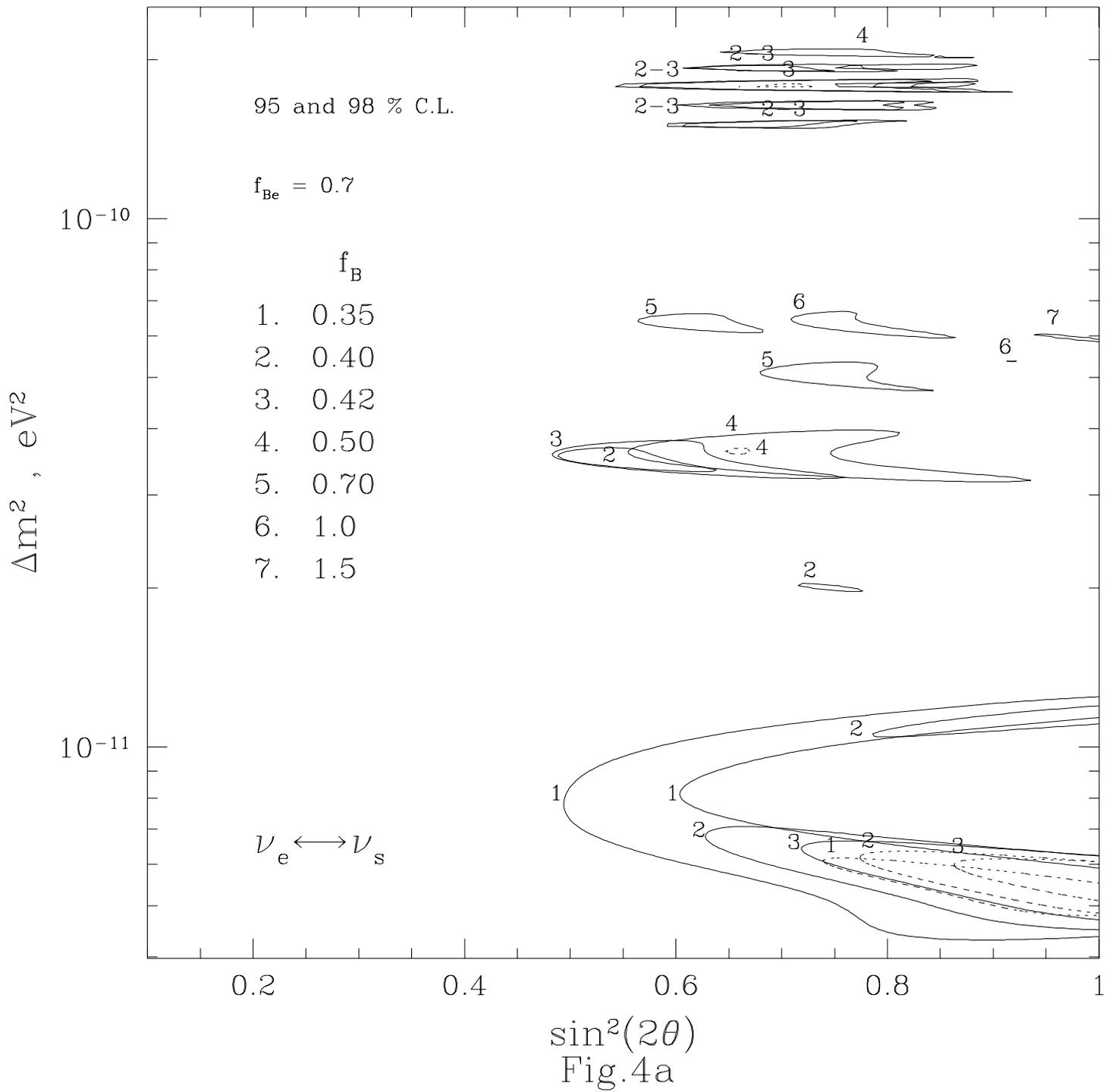
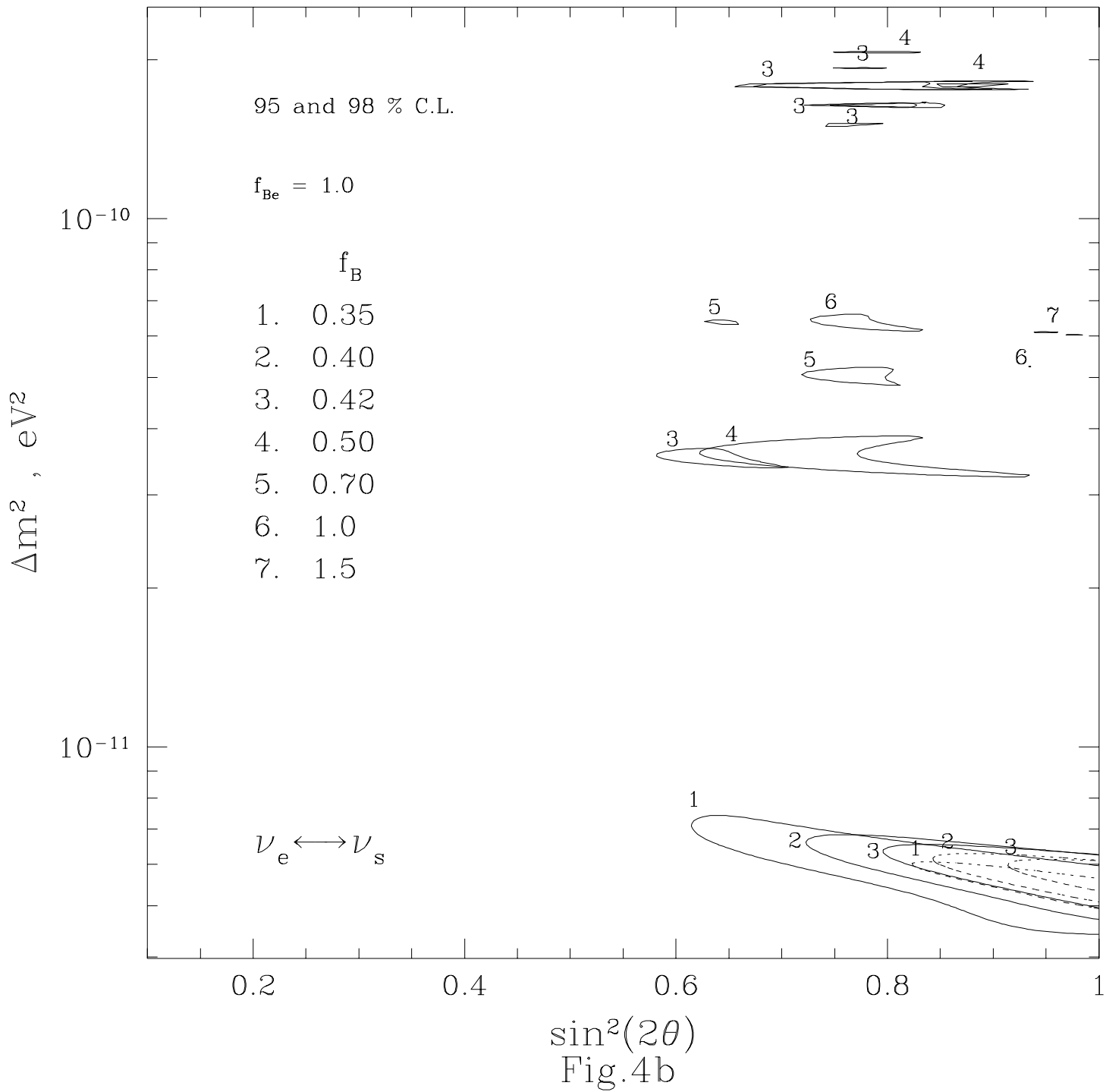
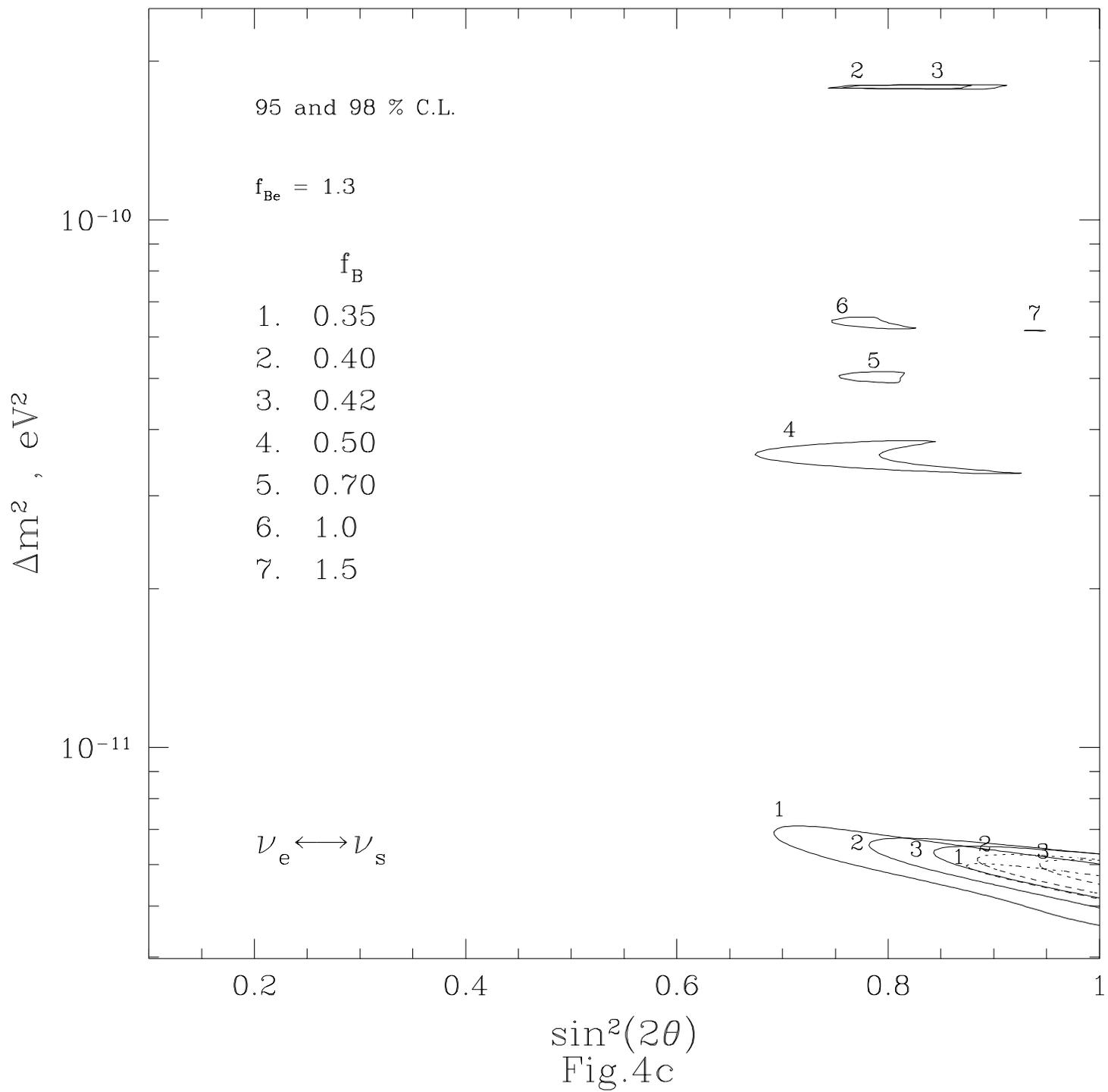
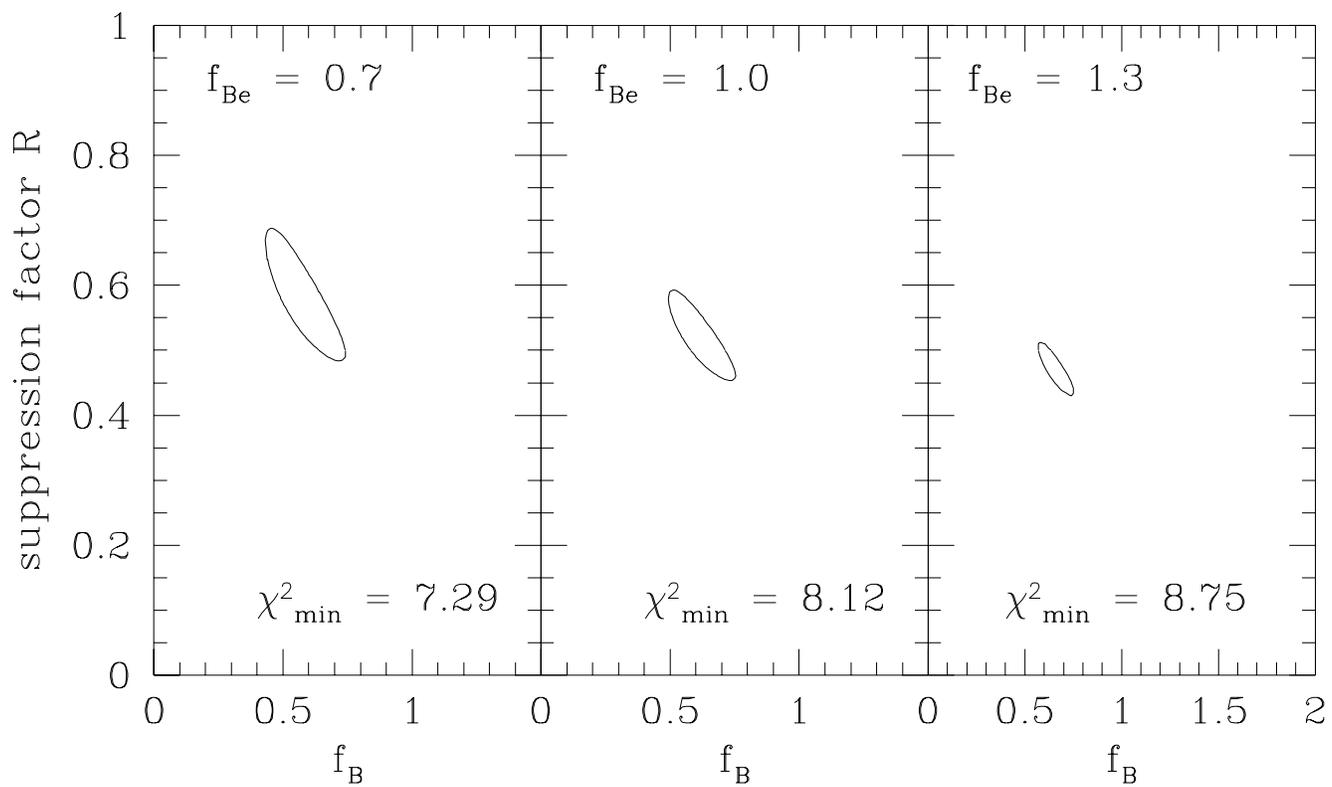
\end{document}